\documentclass[12pt]{article}
\usepackage{amsmath,amsfonts,amsthm,amssymb}
\usepackage{graphicx}
\usepackage{epic,eepic}
\usepackage{times}
\usepackage{multicol}
\usepackage[margin=1in]{geometry}

\newtheorem{theorem}{Theorem}

\usepackage[small]{caption}

\usepackage{hhline}

\newcounter{kevinslistcountertoo}%
\renewenvironment{enumerate}{
\begin{list}{(\roman{kevinslistcountertoo})}
{\usecounter{kevinslistcountertoo}
\setlength{\parsep}{3pt}
\setlength{\labelwidth}{24pt}
\setlength{\itemsep}{3pt}
\setlength{\topsep}{3pt}}}{\end{list}}

\renewenvironment{itemize}{
\begin{list}{-}
{\setlength{\parsep}{3pt}
\setlength{\labelwidth}{24pt}
\setlength{\itemsep}{3pt}
\setlength{\topsep}{3pt}}}{\end{list}}

\newcommand{\E}{{\mathbb E}}
\newcommand{\R}{{\mathbb R}}
\newcommand{\tl}{T_{\rm L}}
\newcommand{\tr}{T_{\rm R}}
\newcommand{\rl}{\rho_{\rm L}}
\newcommand{\rr}{\rho_{\rm R}}

\newcommand{\ttop}{T_{\rm top}}
\newcommand{\tbott}{T_{\rm bottom}}
\newcommand{\rtop}{\rho_{\rm top}}
\newcommand{\rbott}{\rho_{\rm bottom}}

\newcommand{\at}{\alpha_T}
\newcommand{\ar}{\alpha_\rho}
\newcommand{\bt}{\beta_T}
\newcommand{\br}{\beta_\rho}
\newcommand{\dr}{\Delta\rho}
\newcommand{\dt}{\Delta T}

\newcommand{\heading}[1]{\medskip\noindent{\bf {#1}}}

\title{Nonequilibrium Steady States for Certain\\ Hamiltonian Models}

\author{Kevin K. Lin\thanks{Department of Mathematics and Program in
    Applied Mathematics, University of Arizona, AZ 85721, USA.  E-mail:
    klin@math.arizona.edu.  KL was supported in part by NSF Grant
    DMS-0907927.}\ \ \ and\ \ Lai-Sang Young\thanks{Courant Institute of
    Mathematical Sciences, New York University, NY 10012, USA.  E-mail:
    lsy@cims.nyu.edu.  LSY was supported in part by NSF Grant
    DMS-0600974.}}

\date{March 20, 2010}

\begin{document}

\maketitle

\begin{abstract}
  We report the results of a numerical study of nonequilibrium steady
  states for a class of Hamiltonian models.  In these models of coupled
  matter-energy transport, particles exchange energy through collisions
  with pinned-down rotating disks.  In~\cite{ey}, Eckmann and Young
  studied 1D chains and showed that certain simple formulas give
  excellent approximations of energy and particle density profiles.
  Keeping the basic mode of interaction in \cite{ey}, we extend their
  prediction scheme to a number of new settings: 2D systems on different
  lattices, driven by a variety of boundary (heat bath) conditions
  including the use of thermostats.  Particle-conserving models of the
  same type are shown to behave similarly.  The second half of this
  paper examines memory and finite-size effects, which appear to impact
  only minimally the profiles of the models tested in \cite{ey}. We
  demonstrate that these effects can be significant or insignificant
  depending on the {\em local geometry}.  Dynamical mechanisms are
  proposed, and in the case of directional bias in particle trajectories
  due to memory, correction schemes are derived and shown to give
  accurate predictions.
\end{abstract}

\section*{Introduction}

Explaining far-from-equilibrium macroscopic phenomena on the basis of
microscopic dynamical mechanisms is one of the basic problems of
nonequilibrium statistical mechanics.  This paper presents the results
of some numerical studies aimed at shedding light on ``micro-to-macro''
questions for a class of mechanical models introduced in
\cite{rateitschak-klages,llm,ey} as paradigms for transport processes.
Here as in \cite{ey}, the basic idea is that macroscopic quantities such
as energy and density profiles can be deduced -- easily and fairly
accurately -- from certain information on local dynamics provided one
makes a couple of simplifying assumptions.  In this paper, we extend the
results in \cite{ey} to a larger class of models and at the same time
examine more closely the extent to which these assumptions hold.  These
questions lead to a systematic study of how geometry impacts memory and
finite-size effects.

The models studied in this paper describe a coupled transport of matter
and energy.  The local dynamics are purely deterministic and
energy-conserving; that is what we mean by ``Hamiltonian'' in this
paper.  The system is attached to (unequal) stochastic heat baths which
both inject particles into the system and absorb those that leave.
Placed throughout the system, evenly spaced, are rotating disks that are
nailed down at their centers; particles exchange energy with them upon
collisions.  This interaction was first used in
\cite{rateitschak-klages} and \cite{llm}; it takes the place of direct
particle-particle interactions, which are more delicate to control.  The
authors of \cite{ey} introduced the following conceptual simplification:
each rotating disk is enclosed in a well-defined area called a {\it
  cell}, with small enough passageways between cells to ensure
equilibration within each cell. Motivated by this ``decoupling" of
internal cell dynamics and cell-to-cell traffic, they proposed a scheme
that yields simple explicit expressions for approximate energy profiles
and particle densities.  The scheme is based on two assumptions: that
cell-to-cell movement takes essentially the form of an unbiased random
walk, and that within each cell, the system attains local thermal
equilibrium quickly.  They found striking agreement between their
proposed formulas and simulation results, with barely visible errors for
chains consisting of no more than 30 cells.

The aims of this paper are twofold.  One is a straightforward extension
of the results in \cite{ey} to a larger class of models.  Keeping the
basic interaction as well as cell structure in the last paragraph, we
have found the prediction scheme in \cite{ey} -- suitably modified -- to
be very flexible. To give a sense of the type of generalizations, our
simulations are in 2D, but we expect similar results to hold in 3D (with
rotating balls in the place of rotating disks).  In dimensions greater
than one, there are different choices of lattices and boundary
conditions. We show that this scheme is equally valid for rectangular
and hexagonal lattices, and for all of the boundary conditions tested.
The resulting profiles are, of course, different in each case. We permit
even point sources and the use of thermostats to regulate temperature in
different parts of the domain.  These and other features are illustrated
in Examples 1--3 in Sect.~2.  We also extend the results in \cite{ey} in
a different direction, to ``closed'' systems, i.e., systems in which the
total particle number is conserved and system-bath interactions are
limited to energy exchange (Sect.~4).

Our second aim is to understand why memory and finite-size effects are
so mild in the models tested, or, put differently, why a scheme that
ignores these effects can perform so well in profile predictions. Here
we find the key to lie in {\it cell geometry}, referring roughly to the
geometric relation of cell walls and passageways to the rotating disk.
To examine these issues in greater depth, we consider a broader class of
cell designs.  Among the geometries studied here, the one that
represents the greatest departure from the ideas in \cite{ey} (see
above) is used in Examples 5 and 6(A) in Sect.~3.3 (respectively
Figs.~\ref{fig:10}(a) and~\ref{fig:10.5}(a)).  In these examples, walls
between cells are broken down altogether leaving a domain with rotating
disks inside.  To summarize, we find that finite-size and memory effects
can be significant or insignificant.  By understanding the dynamical
mechanisms behind them, we learn to identify ``good" and ``bad"
geometries.  With regard to directional bias in particle trajectories
due to memory, an issue also addressed in \cite{eckmann}, in Sect.~3.1
we offer a correction scheme that is proven to be effective for systems
with ``bad geometry.''

\heading{Stochastic idealizations.}  Alongside the Hamiltonian models,
the authors of \cite{ey} introduced a class of Markov jump processes
intended to be their stochastic idealizations.  Our generalizations in
Sects.~2 and 4 apply also to these stochastic models.  We have run many
tests, and have found -- without exception -- that the simulations
converge easily to predicted values (with, not surprisingly, smaller
errors and faster convergence rates than for their Hamiltonian
counterparts).  While this class of stochastic models, sometimes called
``random-halves models" (see \cite{lin-young}), are interesting in their
own right, we have elected to focus on Hamiltonian models in this paper
because they are both more complex and more physical.

\heading{Related works.}  For reviews of some general physical and
mathematical issues involved in the study of nonequilibrium steady
states and transport processes, see, e.g.,
\cite{bonetto-reybellet,gallavotti-cohen,groot,lepri,ruelle1,ruelle2,spohn-book}.
Much more is known in the context of stochastic models or models with a
significant stochastic component.  Of the considerable literature on
that topic, we cite here a few that are most closely related to the
present
study:~\cite{bertini1,bertini1a,bertini2,bonetto-olla,derrida,derrida1,kipnis-landim,kmp,lepri-mejia,lin-young,olla-varadhan-yau,ravi-lsy}.
The literature on models with purely deterministic internal dynamics is
more sparse.  In addition to the articles already cited,
\cite{balint,bricmont-kupiainen1,collet-eck,collet-eck-carlos,eck-hairer,eck-jacquet,gaspard,hairer-mattingly,li-casati,reybellet1,reybellet2,ruelle3}
are the main ones we know of.  Finally, aside from putting \cite{ey} in
the more general contexts to which it belongs, we have taken some small
steps in this paper to connect geometry and local dynamics to
nonequilibrium properties, a connection that, in general, remains to be
understood.

\section{Review of models and results in \cite{ey}}

Since the present paper is an extension of \cite{ey} with a focus on
Hamiltonian models, we begin with a brief review of the models and  
relevant findings in the Hamiltonian part of \cite{ey}. 

\subsection{Description of models}

\begin{figure}
  \begin{center}
    \includegraphics[scale=0.6]{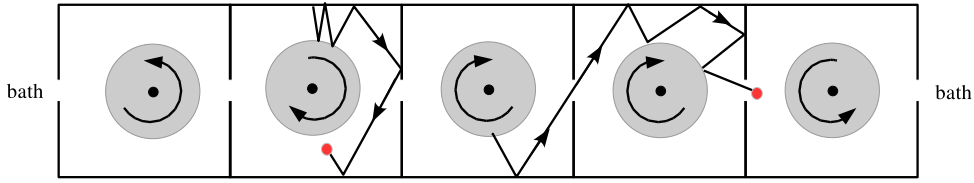}
  \end{center}
  \caption{Illustration of 1D chain and baths.}
  \label{fig:1}
\end{figure}

\heading{Single cell in isolation.}  Let $\Gamma_0\subset\R^2$ be a
bounded, simply connected planar region with piecewise-$C^3$ boundary.
A disk $D_0$ is placed in the interior of $\Gamma_0$; its center is
fixed, and it is allowed to rotate freely about its center. The state of
the disk is given by $(\theta, \omega)$, where $\theta$ is the angle
(relative to a marked reference point) and $\omega$ its angular
velocity.  A number of particles move about in the cell.  The state of a
particle is given by $(x,v)$, where $x\in\Gamma := \Gamma_0\setminus
D_0$ is its position and $v\in\R^2$ its velocity.  Particles fly freely
with constant velocity until they hit either $\partial\Gamma_0$ or
$\partial D_0$ (they do not interact with each other directly).  A
particle that collides with $\partial\Gamma_0$ reflects elastically:
\begin{displaymath}
  v^{\parallel}(t_0^+) = v^{\parallel}(t_0^-)~,\qquad v^{\perp}(t_0^+)
  =~ -v^{\perp}(t_0^-)~,
\end{displaymath}
where $t_0$ is the collision time, $v^{\parallel}$ is the component of
the particle's velocity tangential to $\partial\Gamma_0$, and
$v^{\perp}$ is the normal component.  When a particle hits $\partial D_0$,
the particle swaps its tangential velocity with the disk's angular
velocity:
\begin{align*}
  \omega(t_0^+) = v^{\parallel}(t_0^-)~,&\qquad v^{\parallel}(t_0^+) =
  \omega(t_0^-)~,\\[1ex]
  v^{\perp}(t_0^+) &= ~-v^{\perp}(t_0^-)~.
\end{align*}
We assume that the masses of the particles and the moment of inertia of
the rotating disk are scaled so that the kinetic energy of each particle
is $|v|^2$ and the rotational energy of the disk is $\omega^2$~.  Energy
is conserved in the interaction rule above.

\heading{Single cell coupled to two heat baths.} Suppose now the cell
has two openings, which we identify with subsets $\gamma_L,
\gamma_R\subset\partial\Gamma_0$~. It is useful (though not necessary)
to think of $\Gamma_0$ as having a left-right symmetry, and $\gamma_L$
and $\gamma_R$ as symmetrically placed on the left and right sides of
$\Gamma$ respectively. These two openings are connected to two {\em heat
  baths} that absorb and emit particles.  A particle absorbed by one of
the baths leaves the system forever.  Each bath is characterized by two
parameters, a {\em temperature} $T$ and an {\em injection rate} $\rho$.
A Poisson clock of rate $\rho$ is associated with each bath.  When its
clock rings, the bath places a new particle into the system at a
(uniformly) random position $x\in\gamma$, where $\gamma = \gamma_L$ for
the left bath and $\gamma=\gamma_R$ for the right bath.  The particle is
assigned a random (I.I.D.) velocity $v$ sampled from the distribution
\begin{equation}
  c e^{-\beta |v|^2}|v||\sin\varphi|~dv~,\qquad
  c=\frac{2\beta^{3/2}}{\sqrt\pi}~,
  \label{eq:bath}
\end{equation}
where $\beta = 1/T$ and $\varphi\in[0,\pi]$ is the angle between $v$ and
$\gamma$, measured so that $\varphi=\frac\pi{2}$ points {\em into}
$\Gamma$~. The actions of the two baths are independent.

\heading{1D-chain coupled to two heat baths.} Consider $N$ copies of the cell described
above -- call them $C_1, C_2, \cdots, C_N$ -- each with two 
openings $\gamma_L$ and $\gamma_R$ symmetrically placed. 
We think of these cells as occupying integer sites $1,2,\cdots,N$, 
and for each $i$ identify the right opening $\gamma_R(C_i)$ of the 
$i$th cell with the left opening $\gamma_L(C_{i+1})$ of the $(i+1)$st cell, 
so that particles are able to pass from cell to cell along the chain.
Sites $0$ and $N+1$ are occupied by heat baths which inject
particles into (and absorb particles from) cells $C_1$ and $C_N$
respectively. The rules of injection are as above.

\heading{Remarks.} (i) We note that except for the chain-bath
interaction, which is stochastic in nature, the dynamics within the
chain are purely deterministic and energy-conserving.  It is for this
reason that we refer to these models as Hamiltonian chains.
The local dynamics are, in fact, not symplectic at collisions
  between particles and rotating disks. (ii) Even though particles do
not interact directly with each other, they do interact quite strongly
via the disks, and the overall character of the dynamics is very
strongly nonlinear.

\vspace*{1.5ex}

\heading{Equilibrium distributions.}  The following result characterizes
the Hamiltonian chain in thermal equilibrium with two equal heat baths.

\begin{theorem}[Eckmann-Young \cite{ey}]

  Consider an $N$-chain in contact with two baths, both with
  temperature $T$ and rate $\rho$~:
  \begin{enumerate}

  \item {\bf Single cells.}  For $N=1$, the system preserves a 
  probability measure
    $\mu^{T,\rho}$ characterized by the following properties: The
    number $\kappa$ of particles in the cell is a Poisson random
    variable with mean
    \begin{displaymath}
      2\sqrt\pi \cdot \frac{\mbox{area}(\Gamma)}{|\gamma|} \cdot
      \frac{\rho}{\sqrt{T}}~,
    \end{displaymath}
    where $|\gamma|$ is the length of an opening.  The conditional
    measure of $\mu^{T,\rho}$ on $\Omega_k$, the
state space for a single cell containing $k$ indistinguishable particles,
is given by
    \begin{align*}
      &c_k \exp\Big(-\beta\big(\omega^2 + \sum_{j=1}^k
      |v_j|^2\big)\Big)~d\theta~d\omega\cdot~dx_1\cdots dx_k\cdot dv_1\cdots
      dv_k~,
    \end{align*}
    where $\beta = 1/T$ and $c_k$ is the normalizing constant.

  \item {\bf $N$-chains.}  Let $\mu_i^{T,\rho}$, $i=1,2,\cdots,N$, be $N$
    copies of $\mu^{T,\rho}$.  An $N$-chain in contact with
    two equal baths held at $(T,\rho)$ preserves the product measure
    $\operatorname*{\otimes}_{i=1}^N \mu_i^{T,\rho}$~.

  \end{enumerate}
  \label{prop1}
\end{theorem}

\medskip
\noindent
Here are a few different ways to measure temperature for a cell with distribution $\mu^{T,\rho}$:
\begin{enumerate}

\item The mean disk energy, $\E[\omega^2]$, is $\frac12 T$~.

\item The mean energy per particle in the cell, $\E[|v|^2]$, is $T$.

\item The expected energy of a particle {\em exiting} the cell is $\frac32 T$;
its distribution is given by Eq.~(\ref{eq:bath}).
\end{enumerate}

\heading{Nonequilibrium steady states.}  Suppose a chain is connected to
two unequal heat baths, i.e., $(\tl,\rl)\neq(\tr,\rr)$, where $\tl$ and
$\tr$ are the temperatures of the left and right heat baths
respectively, and $\rl$ and $\rr$ are their respective injection rates.
It was found numerically in \cite{ey} (though not proved) that
irrespective of initial condition, following a transient the system
settles down to a {\em nonequilibrium steady state} determined solely by
the quadruple $(\tl,\rl; \tr,\rr)$.  The product nature of the
equilibrium distribution in Theorem ~\ref{prop1} depends crucially on
having {\em detailed balance} within the system (see \cite{ey}).  In
nonequilibrium steady states, there is no longer detailed balance, and
the invariant measure cannot be explicitly computed. In particular, the
system has spatial correlations, and the steady state distribution is
not of product form.


\subsection{Scheme for predicting nonequilibrium profiles proposed
in \cite{ey}}
\label{sect:1.2}

In this subsection we review the scheme proposed in \cite{ey} for
deducing approximate mean disk energies, temperatures, and other
profiles in chains that are out of equilibrium. This scheme is based on
the following ideas:

\bigskip
\noindent {\it Assumption 1. The system is close to having zero
  directional bias.}  When a particle leaves a cell, let $P_L$ denote
the probability that it exits to the left and $P_R$ the probability that
it exits to the right. By {\bf zero bias}, we mean $P_L= P_R=\frac12$,
regardless of the state of the particle at the time it entered the cell
or the state of the cell while it is there.

\bigskip
\noindent {\it Assumption 2. The system is close to local equilibrium.}
Given $\tl$, $\tr$, $\rl$, and $\rr$ (with $(\tl,\rl)\neq(\tr,\rr)$) and
$x \in (0,1)$, let $\mu_{N,[xN]}$ denote the marginal distribution of
the invariant measure for an $N$-chain at site $[xN]$. We assume there
are numbers $T=T(x)$ and $\rho=\rho(x)$ independent of $N$ such that for
the $N$-chain in question, $\mu_{N,[xN]}$ is close to
$\mu^{T(x),\rho(x)}$ where $\mu^{T,\rho}$ is as in Sect.~2.1.

\bigskip
The rationale for the first assumption is that when the length of the
opening $|\gamma|$ is small, a particle will remain inside each cell it
visits for a long time, eventually losing memory of past events. The
second assumption is based on the idea of {\it local thermal
  equilibrium} ({\bf LTE}) \cite{groot,kipnis-landim}, which, if true
for these models, will imply the convergence of $\mu_{N,[xN]}$ to
$\mu^{T(x),\rho(x)}$ as $N \to \infty$. The idea in \cite{ey} is to
treat the approximations in Assumptions 1 and 2 as though they were
exact, and to derive under these simplifying assumptions mean energy and
particle density profiles.

\heading{Derivation of profiles assuming zero bias and local equilibrium.}
All means below are taken at steady state. For each cell $i$, let
\begin{align*}
  J_i &= \mbox{mean number of particles exiting cell $i$ per unit time,}\\[1ex]
  Q_i &= \mbox{mean total energy exiting cell $i$ per unit time.}
\end{align*}
By $J_i$, we refer here to the total number of particles exiting cell
$i$ via either one of its two exits, and similarly for $Q_i$.  Taking
the approximation in Assumption 1 to be exact, we get, from the
conservation of mass and energy,
$$
J_i = \frac12(J_{i+1} +  J_{i-1})~,\qquad
Q_i = \frac12(Q_{i+1} + Q_{i-1})~,\qquad i=1,2,\cdots,N~,
$$
with boundary conditions
$$
J_0 = 2\rl~,~~~J_{N+1} = 2\rr~,~~~Q_0 = 2\rl\cdot\frac32\tl~,~~~Q_{N+1} = 2\rr\cdot\frac32\tr~.
$$
It follows that the quantities $J_i$ and $Q_i$ linearly interpolate between the
boundary conditions.  As $N\to\infty$, $J_{[xN]}$
and $Q_{[xN]}$ converge (trivially) to
\begin{align*}
  J(x) &= 2\rl(1-x) + 2\rr x~,\\[1ex]
  Q(x) &= 3\rl\tl(1-x) + 3\rr\tr x~
\end{align*}
for all $x\in(0,1)$~.  Returning to the $N$-chain, the local equilibrium
assumption together with Theorem~\ref{prop1} then tells us that for cell
$i$, we have $Q_i = J_i\cdot\frac32 T_i$, leading to the
$$
\mbox{mean temperature profile} \quad
T(x)~~=~~\frac23\cdot\frac{Q(x)}{J(x)}
~~=~~\frac{\rl\tl(1-x) + \rr\tr x}{\rl(1-x) + \rr x}~.
$$ (We think of the $i$th cell as occupying location $x_i=\frac{i}{N+1}$
and the baths as located at $x\in\{0,1\}$.)  Similarly, we find the
\begin{align*}
  \mbox{mean disk energy profile } \ D(x) &:= \lim_{N\to\infty}\E[\omega_{[xN]}^2] ~~=~~
  \frac12T(x)~,\\[2ex]
  \mbox{mean particle density profile } \ \kappa(x) &:= \lim_{N\to\infty}\E[\kappa_{[xN]}] ~~=~~ 2\sqrt\pi\cdot
  \frac{area(\Gamma)}{|\gamma|} \cdot \frac{J(x)}{2\sqrt{T(x)}}~.
\end{align*}
Additional profiles, e.g., total kinetic energy in each cell, are also
easily deduced.

\heading{Remarks.} Mathematically, that the formulas above give
approximate profiles for sufficiently long chains of cells with
sufficiently small openings will follow from the continuity of the
observables $T, D$ and $\kappa$ in the double limit of zero-bias and
infinite volume, the former to be interpreted as $|\gamma| \downarrow
0$.  Such limits are far from trivial. Rigorous results are out of reach
at the present time.

\medskip
In \cite{ey}, simulations were performed for models in which $\Gamma_0$
is given by the area enclosed by four overlapping
disks with small openings introduced, and results were reported 
for 1D-chains with various parameters.  
It was found that even for relatively small $N$, e.g., $N =
30$, the predictions above compare surprisingly well against empirical data.
This suggests that for the models tested, (i) memory effects leading
to directional bias in particle movements are indeed negligible, and
(ii) local equilibrium is, to a good approximation,
achieved at relatively small $N$, 
i.e., finite size effects are insignificant.

\section{Generalization to larger class of models}
\label{sect2}

In this section, we generalize the results of \cite{ey} in a variety of
directions, including:

\smallskip
\begin{enumerate}

\item[-] {\em Higher dimensions.}  We focus on 2D models; phenomena
  in higher dimensions are expected to be similar.

\item[-] {\em Lattice types.}  In two or more dimensions, there are
  different lattice types.  We focus here on rectangular and hexagonal
  geometries.

\item[-] {\em Boundary conditions.}  
Dirichlet, periodic, and reflecting boundary conditions are considered.
We show that our scheme applies even to point sources. 

\item[-] {\em Thermostats.} By this we refer to the use of external means
to maintain or regulate the temperature of various parts of the system.

\end{enumerate}

\smallskip
\noindent
Instead of giving a formal treatment of each of these generalizations,
we will present in Sects.~2.1--2.3 below three examples that
systematically illustrate the various effects.

\heading{General setup and issues.}  The following are features common
to all the models to be presented.  We first discuss them so we can
focus on the specific points of interest within each example.

\begin{figure}
  \begin{center}
    \includegraphics[scale=0.6]{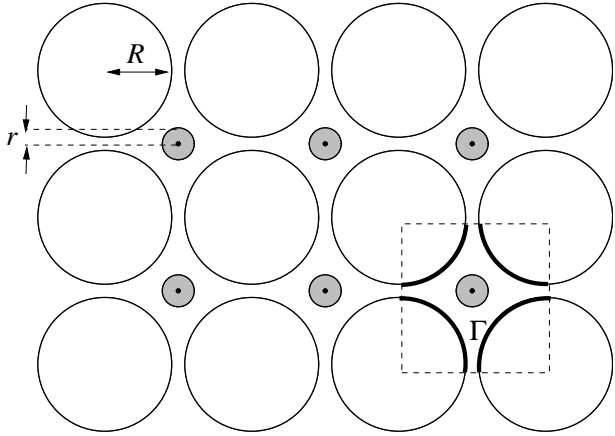}
  \end{center}
  \caption{Illustration of rectangular lattice system.  Here, the system
    consists of a $4\times 3$ array of fixed disks of radius $R$, defining
    among them a total of 6 cells. At the center of each cell is a rotating
    disk (shaded with centers marked) of radius $r$.  A
    cell $\Gamma$ is identified.}
  \label{fig:2}
\end{figure}

\smallskip
\noindent (1) {\em Basic configurations.} We start with a
lattice of disjoint {\em fixed} (non-rotating) {\em
  disks} in ${\mathbb R}^2$.  Fig.~\ref{fig:2} shows a rectangular
array; a hexagonal array is shown in Fig.~\ref{fig:5}.  Particles move
in the space between these disks, bouncing off them in specular
collisions. The packing is relatively dense, with narrow channels
between adjacent disks to allow the passage of particles. The area in
${\mathbb R}^2$ between this array of fixed disks is partitioned into
identical {\em cells} whose walls are circular arcs (pieces of
boundaries of the fixed disks) alternating with openings called {\em
  gaps}; see Fig. 2. At the center of each cell is a {\em rotating
  disk}, with which particles exchange energy following the rules in
Sect.~1.1. Disk energy will always refer to the kinetic energy of a
rotating disk.

  \smallskip
  \noindent (2) {\em Mixing rates and geometry.}  We use $R$ and $r$ to
  denote respectively the radii of the fixed and rotating disks. These
  two numbers together with the distance between the centers of the
  fixed disks will affect the number of collisions a particle
  experiences before it finds its way out of a cell.  More collisions
  will naturally lead to better equilibration within cells.  Too many
  such collisions, on the other hand, will delay mixing on a more global
  scale and slow down convergence to steady state.  Information on the
  expected number of collisions can be deduced from Theorem 1.  For
  example, on each visit to a cell, the number of collisions between a
  particle and the rotating disk is, on average, $(2\pi r)\slash
  (\mbox{total length of all gaps})$.

\smallskip
\noindent (3) {\em Continuum limits.}  Our domains of interest are
bounded. Formally, we fix a region $\Lambda \subset {\mathbb R}^2$ with
piecewise smooth boundary, and consider a sequence of homothetically
magnified images $\Lambda_n$ of $\Lambda$. At stage $n$, the fixed disks
in the system are those that meet $\Lambda_n$. To obtain continuum
limits of profiles for disk energy, particle counts, etc., we map
$\Lambda_n$ back to $\Lambda$ (so that the profiles for different $n$
are all defined on $\Lambda$) and take $n\to\infty$.

\smallskip
\noindent (4) {\em Predicted vs. simulated results.}  In each of the
examples below, we will report the percentage discrepancy between the
predicted and simulated values of energy, particle count, etc.  
Our predictions are based on exact versions of the two assumptions in
Sect.~\ref{sect:1.2}, which cannot be true for finite chains with finite
gap sizes, and the simulated values given are computed
empirically for finite -- and only moderately large -- systems.

\heading{Remark on numerical simulations.}  
All numerical results in this paper are computed by direct Monte Carlo
computations, i.e., expectations are estimated by ergodic averages based
on long-time simulations.  We prepare the system by sampling the
equilibrium distribution at a uniform temperature $T_{init}$, where
$T_{init}$ is usually chosen to be the {\em maximum} of the boundary
temperatures, and evolve the system for $10^{10}$ events (an event is
a single particle-disk collision, bath injection, or particle exit).
Statistical errors are quantified by a standard batch means method.
Typical relative errors for the simulations in this section are $\leq 2\%$.
For example, for the results in Fig.~\ref{fig:3} below, the (absolute)
standard error for the $D$ profile is estimated to be $\leq 0.05$ and
the relative error is $\leq 0.4\%$ at the 95\% confidence level.

\subsection{Example 1. Rectangular lattice, Dirichlet  boundary
conditions}

Our first example is a 2D rectangular array on a rectangular-shaped
 domain with
Dirichlet boundary conditions (to be explained).  Fixed disks of radius
$R\in(0,1)$ are placed at {\em even} lattice points
$\{0,2,\cdots,2M\}\times\{0,2,\cdots,2N\}~.$ We think of $R$ as being
close to $1$, so that each square configuration of 4 fixed disks bounds
an open cell with 4 narrow gaps, with gap size $2(1-R)$.
In the center of each cell, i.e., at
odd lattice points $\in\{1,\cdots,2M-1\}\times\{1,\cdots,2N-1\}$, we
place a small rotating disk of radius $r$. 
Heat baths are viewed as occupying the strip of
cells just outside the system, injecting and removing particles through
the associated openings.  For simplicity, we consider only
configurations where all baths along the same edge have the same
parameters $(T,\rho)$. For example, if the bath along the left edge has
parameters $(T_L, \rho_L)$, then particles with energies distributed
according to Eq.~(\ref{eq:bath}) with $T=T_L$ are injected at rate
$\rho_L$ into each cell in the leftmost column, through its opening on
the left.  We refer to the boundary condition described above as a
``Dirichlet'' boundary condition.

\heading{Profile prediction.}  The scheme reviewed in
Sect.~\ref{sect:1.2} is easily generalized to the present setting.
Formulating Assumption 1 in 2D is straightforward: it is equivalent to
requiring that the probability of exiting any of the 4 openings to be
$\approx\frac14$.  The precise meaning of LTE here is as follows: let
$\Lambda = [0,a] \times [0,b] \subset {\mathbb R}^2$ be the fixed domain
on which the continuum limit will be taken, and set $M=[na]$ and
$N=[nb]$ for positive integers $n$.  Then for every
$(x,y)\in(0,a)\times(0,b)$, the marginal distribution of the cell at
site $([nx],[ny])$ tends to $\mu^{T,\rho}$ for some $T=T(x,y)$ and
$\rho=\rho(x,y)$ as $n \to \infty$.

Returning to the finite system described in the first paragraph,
let $J_{k,\ell}$ and $Q_{k,\ell}$ denote the particle and energy exit
rate from the $(k,\ell)$th cell. The particle balance equation is the
discrete Laplace equation
$$
J_{k,\ell} = \frac14\big(J_{k+1,\ell} + J_{k-1,\ell} + J_{k,\ell+1} + J_{k,\ell-1}\big)~,
$$ with boundary conditions $J = 4\rl$ on the left edge, $J = 4\rr$ on
the right edge, etc.  The $Q_{k,\ell}$ are described by the discrete
Laplace equation as well, with boundary
conditions $Q= 4\rl \cdot \frac32 \tl$ on the left, 
$Q= 4\rr \cdot \frac32 \tr$ on the right, and so on.  In the
continuum limit, these equations tend to the Laplace equation with the
given boundary values.  Solutions for $J$ and $Q$ are then used to
compute the mean disk energy and mean particle count through
\begin{align*}
  D(x,y) &=~~ \frac12T(x,y)~~=~~\frac13\cdot\frac{Q(x,y)}{J(x,y)}~,\\[3ex]
  \kappa(x,y) &= 2\sqrt\pi\cdot
  \frac{4-\pi(R^2+ r^2)}{2(1-R)} \cdot
  \frac{J(x,y)}{4\sqrt{T(x,y)}}~.
\end{align*}
These formulas are virtually identical to those for 1D-chains given in
Sect.~1.2, differing only in some constants, and can be deduced the same
way. The differences in constants are due to the fact that each cell
here is in contact with 4 (rather than 2) cells/baths.

\begin{figure}
  \begin{center}
    \begin{small}
      \begin{tabular}{cc}
        \includegraphics[scale=0.85]{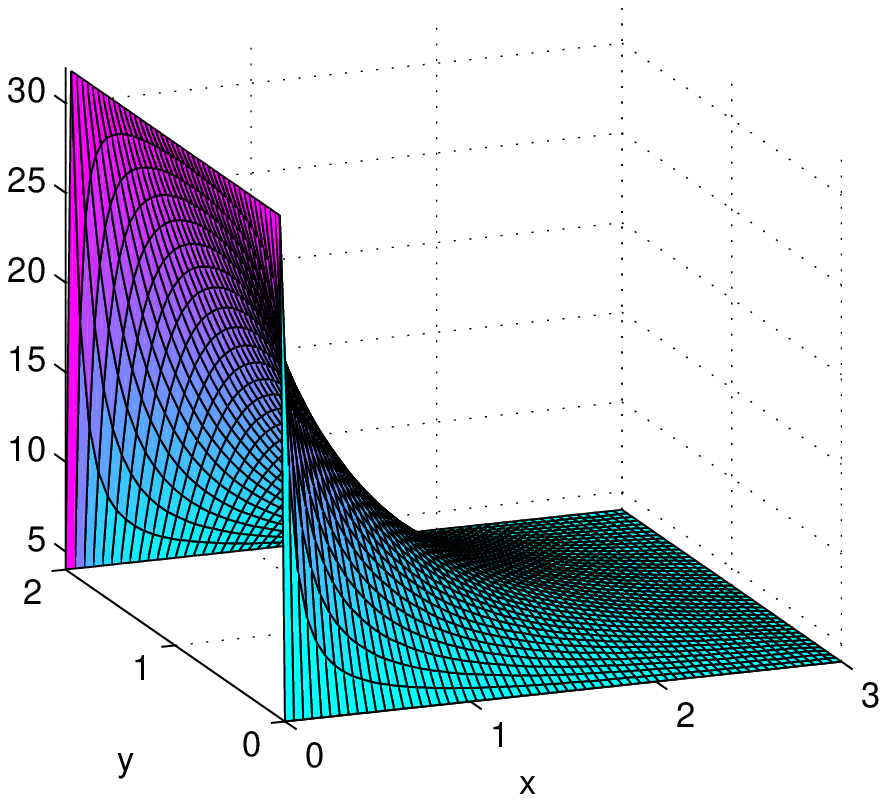} &\includegraphics[scale=0.85]{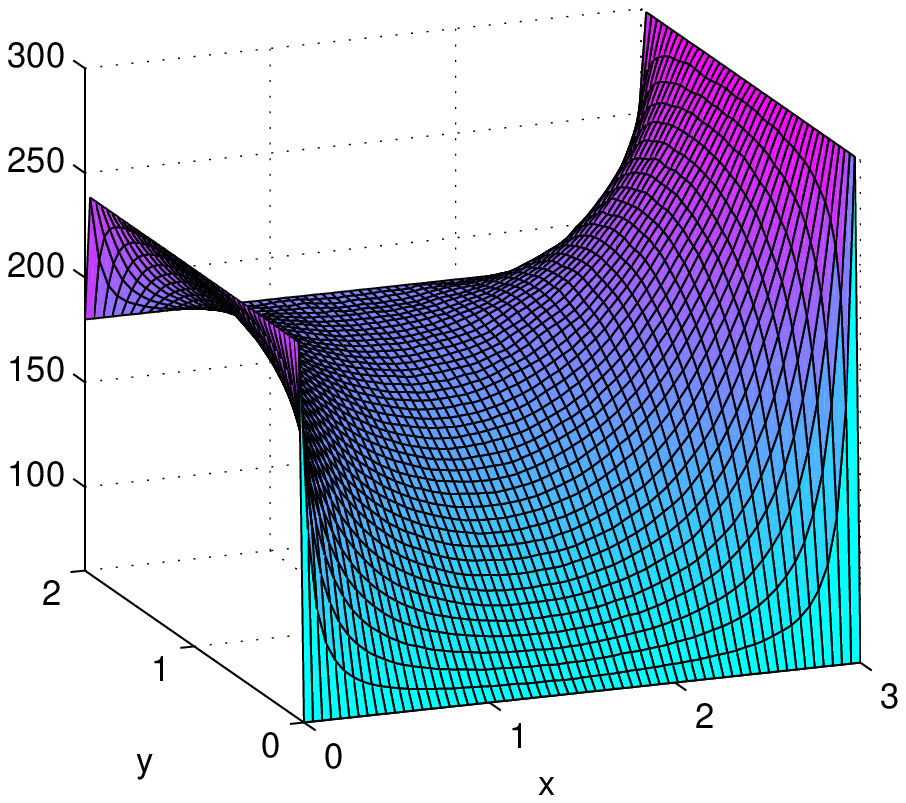} \\
        (a) $J(x,y)$ & (b) $Q(x,y)$ \\[3ex]
        \includegraphics[scale=0.85]{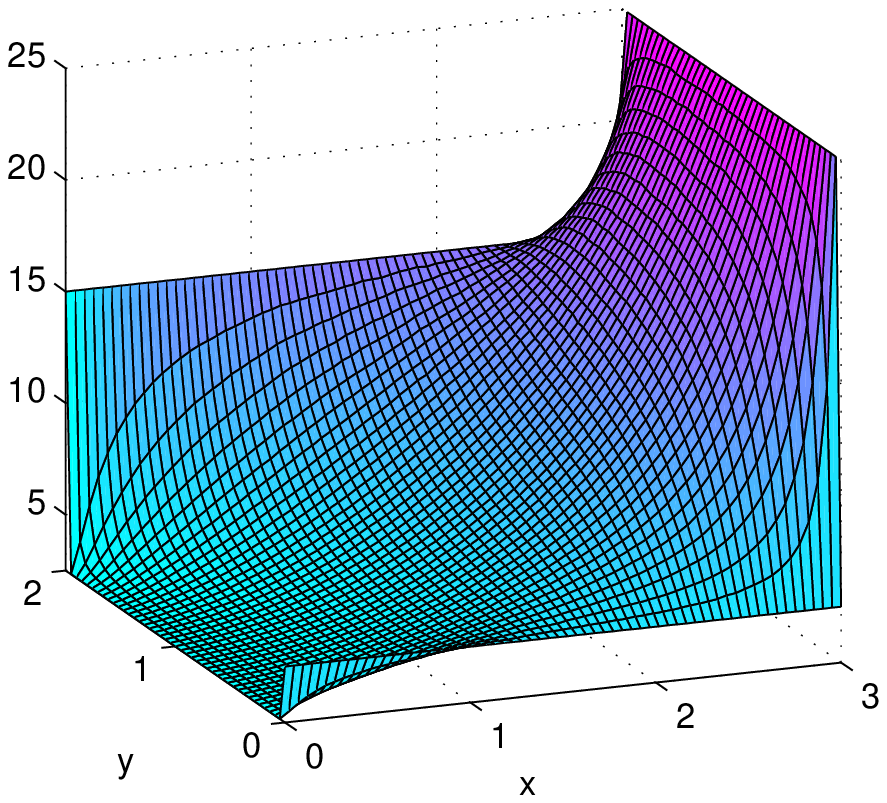} &\includegraphics[scale=0.85]{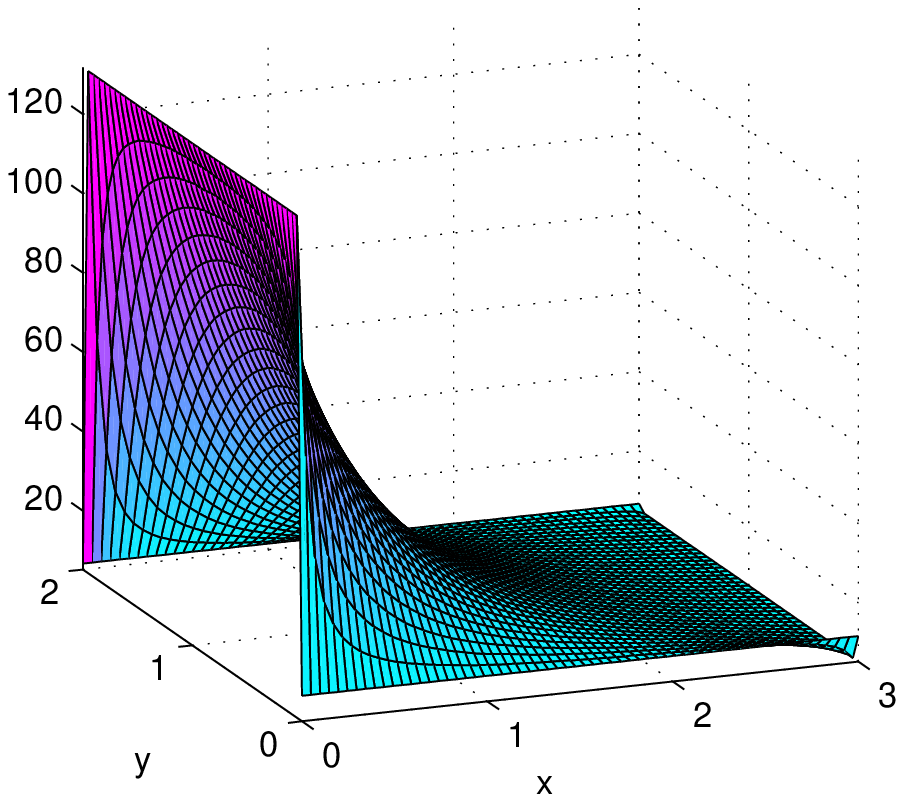} \\
        (c) Disk energy $D(x,y) = \frac12 T(x,y)$ & (d) Particle count $\kappa(x,y)$ \\
      \end{tabular}\\[6ex]
    \end{small}
  \end{center}

  \caption{Simulation results for the rectangular lattice with Dirichlet
    boundary conditions.  Bath parameters are as given in the text.  The
    lattice contains $60\times 41$ rotating disks, which are viewed as
    located on a regular grid within the domain $[0,3]\times[0,2]$.  The
    disk radii are $R=0.95$ and $r=0.2$.  Simulation values are plotted
    over interior grid points, and boundary values are added to ease
    visualization.  For all profiles, the discrepancy between computed
    and predicted profiles are 1-1.2\% for $(x,y)$ away from
    ``L''-shaped regions near the corners.}
  
  \label{fig:3}
\end{figure}

\heading{Simulation results.}  We now compare the predicted profiles to
the results of direct simulations.  Fig.~\ref{fig:3} shows some
simulation results using the following boundary conditions:
\begin{small}
  \begin{center}
    \begin{tabular}{rcl}
      & $(\ttop,\rtop)=(30,1)$ &\\[2ex]
      $(\tl,\rl)=(5,8)$ & \fbox{\parbox{2in}{
          \begin{center}
            \vspace{2ex}
            $60\times 41$ cells
            \vspace{2ex}
      \end{center}}} & $(\tr,\rr)=(50,1)$\\[6ex]
      & $(\tbott,\rbott)=(10,1)$ &\\
    \end{tabular}
  \end{center}
\end{small}
Fixed disks have radius $R=0.95$ and rotating disks $r=0.2$.  These
numbers ensure that on average, each particle experiences $\approx 18$
collisions and hits the rotating disk $\approx 3$ times on each pass
through a cell.

Fig.~\ref{fig:3} shows profiles for the jump rate $J$, the energy exit
rate $Q$, the mean disk energy $D$, and the particle count $\kappa$.
Boundary values have been filled in for ease of visualization.  We have
found that the discrepancy between simulation results and predicted
profiles are greatest in  the ``L-shaped'' regions at corners along
the boundary of the domain. Away from these regions, the discrepancy 
is $\leq 1.5\%$~. That the discrepancy is larger here is clearly 
due to discontinuous
boundary values and finite-size effects.

Before leaving this example, let us check our intuition   
against the plotted profiles.  Observe first that panels (a) and (c), {\it i.e.},
plots of jump rates $J$ and disk energies $D$, reflect the boundary 
conditions in a straightforward way. The saddle-shaped profile of $Q$ 
(panel (b)) is also easy to explain: it is because along the 4 edges 
clockwise from left, $Q = 4\rho \times \frac32 T$ has ratios 4~:~3~:~5~:~1~.
A factor that influences particle count is that 
all else being equal, particles tend to gather in
low-temperature regions.  The reason is that in a steady state, the
total particle flux into and out of any region must be equal.  At the
lower temperature end, particles move slowly, hence more of them are
needed to produce the same flux.  For the boundary conditions considered here,
the accumulation of particles on the left due to colder temperatures is
exaggerated by the higher injection rate on the same side. This has
resulted in the huge spread for $\kappa$ in panel (d), from $> 120$ 
on the left to $<5$ on the right.
If we change the boundary conditions to $\rho_L < \rho_R$, the profile
for $\kappa$ would be quite a bit flatter.  This is both predicted and
confirmed in simulations.

\heading{Non-Dirichlet boundary conditions.}  
We have obtained similar simulation results
for systems with periodic and reflecting boundary conditions in the
vertical direction.



\subsection{Example 2. System cooled by thermostats}

The main novel feature of this example is the use of thermostats in the
present context.\footnote{See, e.g.,
  \cite{chernov,evans-morriss,gallavotti,gallavotti-presutti3,hoover,mejia-rondoni,ruelle0}
  for more examples and thorough discussions of thermostats in
  nonequilibrium statistical mechanics.}  A second feature of note is
the use of exit-only boundaries, which for convenience we will refer to
as {\em open boundaries}.

\heading{Thermostats.}  For our purposes, a {\em thermostat} is a disk
with the property that whenever a particle with velocity $v$ hits it,
the tangential component $v^{\parallel}$ of the particle's velocity is
instantaneously set to a random value $\omega$ sampled from the
distribution
\begin{displaymath}
  \omega \sim c e^{-\beta_* \omega^2}~d\omega~,\qquad
  c=\sqrt{\frac{\beta_*}{\pi}}~,
\end{displaymath}
where $T_* = 1/\beta_*$ is the thermostat temperature.  The normal
component of $v$ is updated according to the usual rule: $v^{\perp}(t^+)
= -v^{\perp}(t^-)~.$ Thus, a particle that hits the thermostated disk
sufficiently many times would acquire the energy distribution for
temperature $T_*$.  A cell is thermostated if it contains a thermostat
in place of a rotating disk.

\medskip
We consider in this example a {\em cooling system on a finite
  cylindrical domain} with an {\em open boundary} at one end. More
precisely, the domain is defined by a rectangular array of fixed disks
with a periodic boundary condition in the vertical direction. The right
end of the cylinder is put in contact with heat baths, with which it has
the same type of interaction as in the previous example. The left end of
the cylinder has an open boundary, meaning particles can leave but no
new particles enter through it. This configuration leads to a particle
flux across the system, from right to left. As the relatively hot
particles move through the system, they are cooled (either directly or
indirectly) by a row of thermostated cells. For simplicity, we assume
the thermostated cells occupy exactly one row which extends the full
length of the cylinder, and all thermostats 
in these cells are set to
a fixed temperature $T_*$ considerably lower than that of the heat bath.

\heading{Profile prediction.}  To obtain predictions for the various
profiles, we proceed as follows:  

First, one solves for $J$ as before, the only difference being that the
boundary conditions for $J$ are periodic in the $y$ direction.
The open boundary for $x=0$ corresponds to setting $J(0,y) \equiv 0$ for
all $y$. 
Solving the boundary value problem readily yields $J$.

To find $Q$, the left and right boundaries are treated
in a similar fashion as for $J$.  To handle thermostats, we assume
that the array is mapped to a domain $[0,a] \times [0,b]$ with 
$\{y=0\}$ and $\{y=b\}$ identified, and that the thermostats are 
at $y=0$. We then set, for
each $(x,y)$ with $y\in\{0,b\}$, $Q(x,y)=\frac32 J(x,y)T_*$~, where
$J(x,y)$ is as in the last paragraph.  This is again a boundary value
problem, which can be solved to give $Q$.  

Once $J$ and $Q$ have been computed, profiles for $D$ and $\kappa$ are given by
the same formulas as in the previous example.

\medskip
We point out two reasons why one may expect greater discrepancies
between predicted and simulated results in the present model than in the
Dirichlet case: (1) In setting the ``boundary values" for $Q$ along the
row of thermostated cells, we have assumed that the outgoing particles
have energy $\frac32 T_*$. In reality, a highly-energetic particle
that enters a thermostated cell is likely to leave it with somewhat more
energy than $\frac32 T_*$~: it interacts with the rotating disk only a
finite number of times, keeping the normal component of its velocity
each time. (2) At the open boundary, $J$ must necessarily go to zero, as
must therefore particle count.
Since a healthy number of particles is needed in each cell to ensure
proper mixing, the situation near this boundary can be
anomalous. 

Also, with few particles, a practical concern is inadequate updating
during the course of the simulation.

\begin{figure}

  \begin{center}
    \begin{small}
      \begin{tabular}{cp{4in}}
        \includegraphics[scale=0.85]{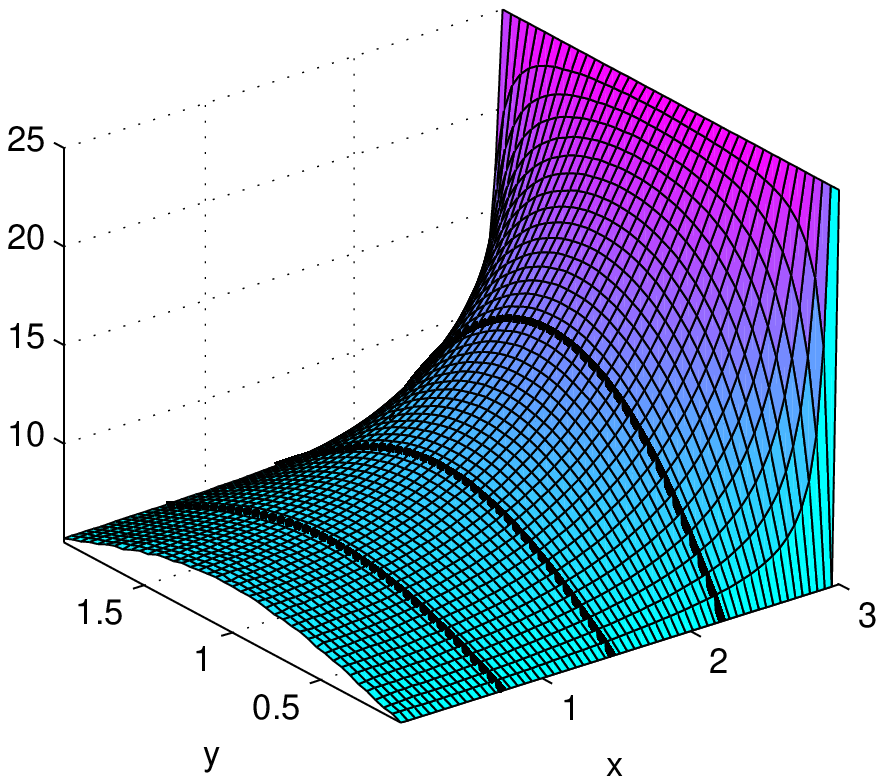} &
        \includegraphics[scale=0.85]{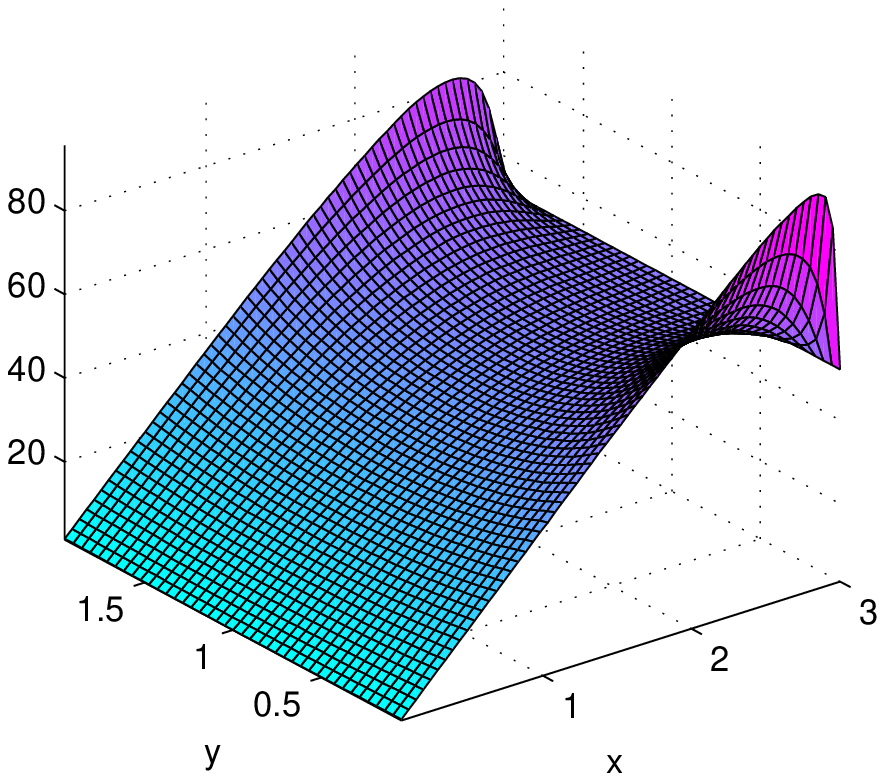} \\
        (a) Mean disk energy $D(x,y)$ &
        (b) Mean particle count $\kappa(x,y)$\\[3ex]
        \includegraphics[scale=0.7,bb=23 128 267 242]{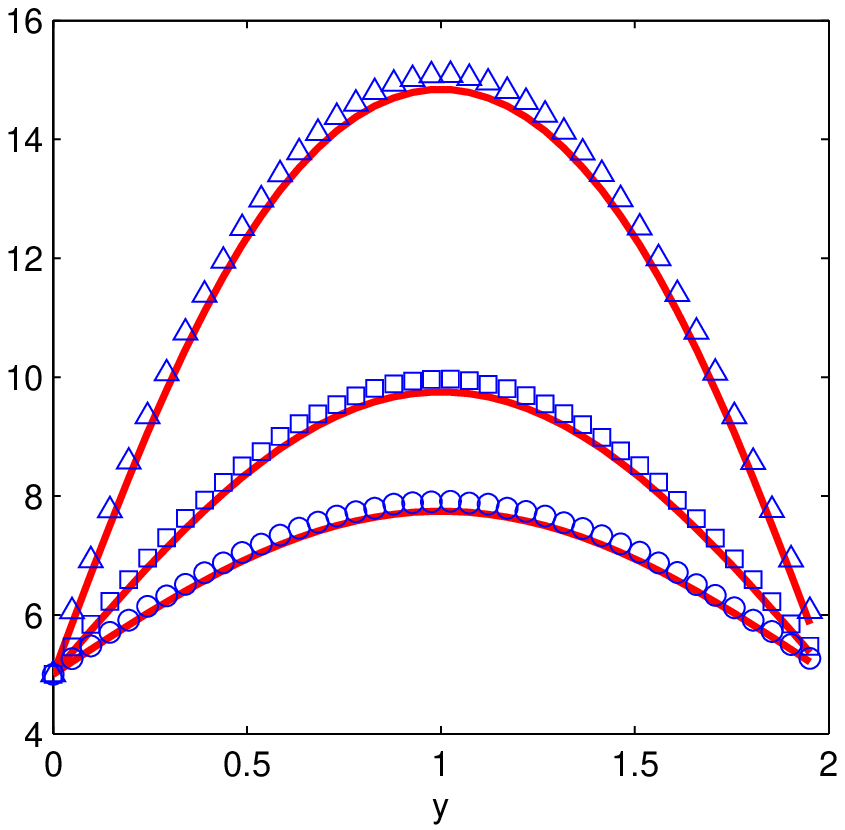} &
        \resizebox{3in}{1.4in}{\includegraphics*[bb=0in 22 3in 2in]{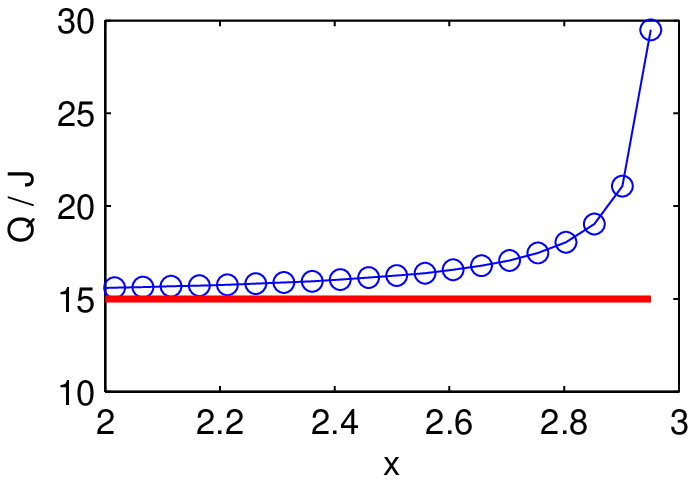}}
        \hspace*{-0.1in}\resizebox{3.1in}{1.4in}{\includegraphics[bb=0in 0in 3in 2in]{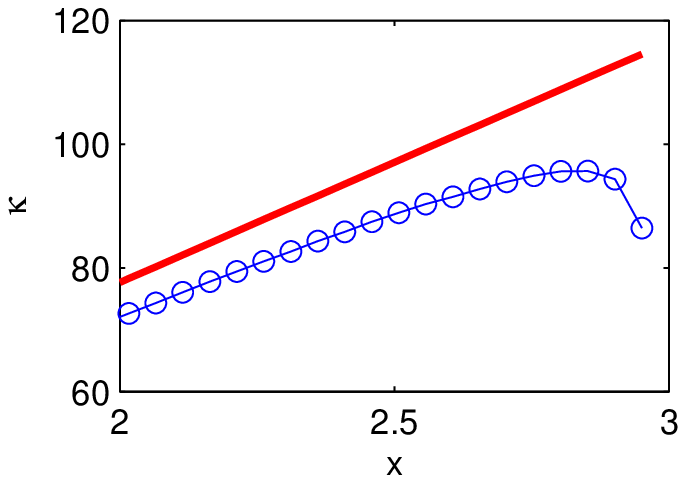}}\\
        (c) Slices of $D$ at $x=0.75, 1.5, 2.25$ &
        (d) Slices of $Q/J$ and $\kappa$ at $y=0$\\[6ex]
      \end{tabular}
    \end{small}
  \end{center}

  \caption{Simulation results for cylindrical array with ``cooling''
    boundary conditions.  The array contains $60\times 41$ rotating
    disks, mapped to grid points inside $[0,3]\times[0,2]$.  Mean disk
    energy and particle count profiles are shown in (a) and (b).  The
    discrepancy away from edges is about 1-4\% in (a) and $\leq 7\%$ in
    (b).  In (c), $D(x,y)$ is plotted against $y$ for $x=0.75$
    ($\circ$), $x=1.5$ ($\square$), and $x=2.25$ ($\Delta$).  The two
    panels in (d) show plots of mean energy per exiting particle $Q/J$
    and particle count $\kappa$ along the row of thermostated cells at
    $y=0$.  Parameters are as described in the text.}
  \label{fig:4}
\end{figure}

\heading{Simulation results.}  Fig.~\ref{fig:4} shows simulation results
for the following boundary values:
\begin{center}
\begin{small}
  \vspace*{2ex}
\setlength{\unitlength}{0.00083333in}
\begingroup\makeatletter\ifx\SetFigFont\undefined%
\gdef\SetFigFont#1#2#3#4#5{%
  \reset@font\fontsize{#1}{#2pt}%
  \fontfamily{#3}\fontseries{#4}\fontshape{#5}%
  \selectfont}%
\fi\endgroup%
{\renewcommand{\dashlinestretch}{30}
\begin{picture}(6453,2150)(0,-10)
\put(4740,333){\ellipse{134}{134}}
\path(490,453)(700,453)(700,220)
	(490,220)(490,453)
\put(797,333){\ellipse{134}{134}}
\path(692,453)(902,453)(902,220)
	(692,220)(692,453)
\put(1007,333){\ellipse{134}{134}}
\path(902,453)(1112,453)(1112,220)
	(902,220)(902,453)
\put(1217,333){\ellipse{134}{134}}
\path(1112,453)(1322,453)(1322,220)
	(1112,220)(1112,453)
\put(1427,333){\ellipse{134}{134}}
\path(1322,453)(1532,453)(1532,220)
	(1322,220)(1322,453)
\put(1637,333){\ellipse{134}{134}}
\path(1532,453)(1742,453)(1742,220)
	(1532,220)(1532,453)
\put(1840,333){\ellipse{134}{134}}
\path(1735,453)(1945,453)(1945,220)
	(1735,220)(1735,453)
\put(2050,333){\ellipse{134}{134}}
\path(1945,453)(2155,453)(2155,220)
	(1945,220)(1945,453)
\put(2252,333){\ellipse{134}{134}}
\path(2147,453)(2357,453)(2357,220)
	(2147,220)(2147,453)
\put(2463,333){\ellipse{134}{134}}
\path(2358,453)(2568,453)(2568,220)
	(2358,220)(2358,453)
\put(2665,333){\ellipse{134}{134}}
\path(2560,453)(2770,453)(2770,220)
	(2560,220)(2560,453)
\put(2875,333){\ellipse{134}{134}}
\path(2770,453)(2980,453)(2980,220)
	(2770,220)(2770,453)
\put(3085,333){\ellipse{134}{134}}
\path(2980,453)(3190,453)(3190,220)
	(2980,220)(2980,453)
\put(3287,333){\ellipse{134}{134}}
\path(3182,453)(3392,453)(3392,220)
	(3182,220)(3182,453)
\put(3497,333){\ellipse{134}{134}}
\path(3392,453)(3602,453)(3602,220)
	(3392,220)(3392,453)
\put(3700,333){\ellipse{134}{134}}
\path(3595,453)(3805,453)(3805,220)
	(3595,220)(3595,453)
\put(3902,333){\ellipse{134}{134}}
\path(3797,453)(4007,453)(4007,220)
	(3797,220)(3797,453)
\put(4105,333){\ellipse{134}{134}}
\path(4000,453)(4210,453)(4210,220)
	(4000,220)(4000,453)
\put(4315,333){\ellipse{134}{134}}
\path(4210,453)(4420,453)(4420,220)
	(4210,220)(4210,453)
\put(4530,333){\ellipse{134}{134}}
\path(4425,453)(4635,453)(4635,220)
	(4425,220)(4425,453)
\put(595,333){\ellipse{134}{134}}
\path(4635,453)(4845,453)(4845,220)
	(4635,220)(4635,453)
\put(4950,333){\ellipse{134}{134}}
\path(4845,453)(5055,453)(5055,220)
	(4845,220)(4845,453)
\path(490,1900)(490,220)
\path(5055,1900)(5055,210)
\put(2255,0){\makebox(0,0)[lb]{{Periodic}}}
\put(1570,565){\makebox(0,0)[lb]{{Row 1 disks: Thermstated at $T_*=10$}}}
\put(2255,2000){\makebox(0,0)[lb]{{Periodic}}}
\put(1650,1145){\makebox(0,0)[lb]{{Total system size: $60\times 41$ cells}}}
\put(5130,1035){\makebox(0,0)[lb]{{$(\tr,\rr)=(50,10)$}}}
\put(0,1070){\makebox(0,0)[lb]{{Open}}}
\dashline{60.000}(490,1900)(5035,1900)
\end{picture}
}

  \vspace*{2ex}
\end{small}
\end{center}
Again, we use $R=0.95$ and $r=0.2$.  Panel (a) shows the mean disk
energy $D$, which is higher on the right side due to bath
injections and lower along the top and bottom edges due to the cooling
effects 
of the thermostats.  Panel (b) shows the particle counts.  In addition to the
cells near the baths, particles accumulate near the top and bottom
boundaries due to the lower temperatures there.  Panel (c) compares
directly simulation data and predicted values for some
cross-sections of the $D$ profile, and panel (d) examines the situation
along $y=0$ (where the thermostats are) for $x \in [2,3]$,
i.e., near the bath. As anticipated, the temperature here is higher 
than its predicted values, with the discrepancy rising sharply as 
$x$ tends to $3$. 
The particle deficit is then consistent with 
the fact that exit rates per particle  are higher than predicted.

The effects highlighted in panel (d), however, seem confined to the
thermostated cells.  Away from them, the discrepancy between predicted
and simulation profiles are about 1-4\% for the mean disk energy and
$\leq 7\%$ for the particle count.
The latter shows larger discrepancies near the open (left) end of the
domain, due presumably to $\kappa\approx 0$ near there.

\subsection{Example 3. Hexagonal array with point sources}

\begin{figure}
  \begin{center}
    \includegraphics[scale=0.7,angle=90,bb=0in 0in 2.7061742934376296in 4.0in]{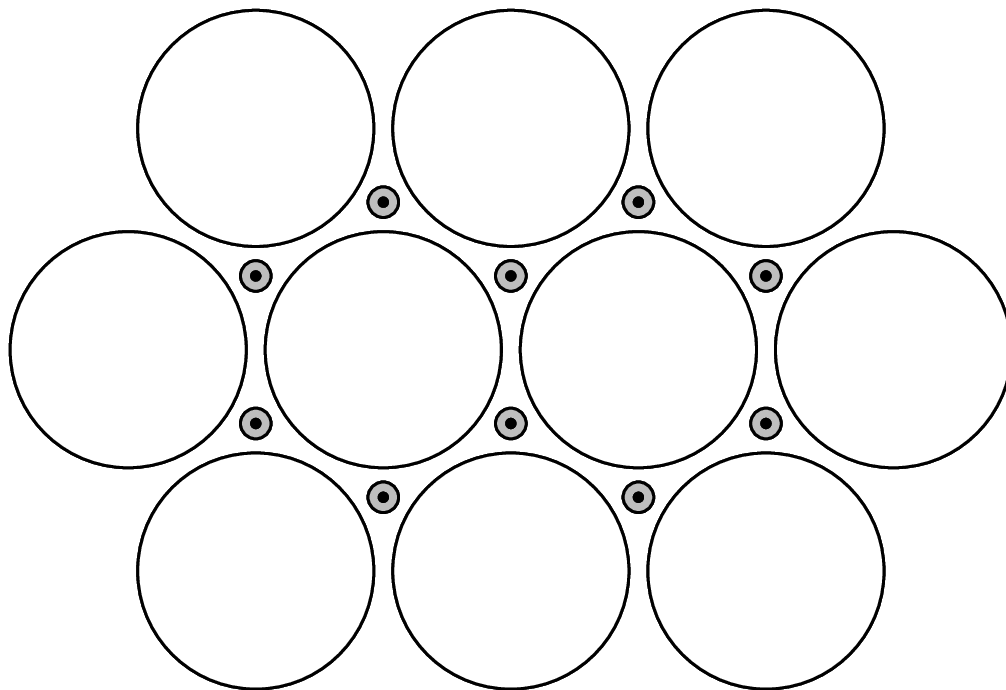}%
  \end{center}
  \caption{A patch of hexagonal lattice.}
  \label{fig:5}
\end{figure}

Our third example is a hexagonal array with point sources and open
boundaries. The array we use consists of fixed disks of radius $R$
placed on the vertices of a triangular mesh; each edge of the mesh has
length 1.  Each triangular array of 3 fixed disks bounds a cell.  At the
center of each cell, we place a small rotating disk of radius $r$. See
Fig.~\ref{fig:5}. The system has open boundaries with no bath
  injections. Instead, it is driven by a finite number of {\em point
  sources} which pump particles directly into the interior of the
array. We say a cell contains a point source with parameters $(T_*,
\rho_*)$ if instead of being an ordinary cell containing a rotating disk, 
it emits particles having the energy distribution in (\ref{eq:bath}) with 
$T=T_*$ at a rate of $\rho_*$~  irrespective of what enters the cell.  
These particles are injected into each of the 3 neighboring cells with 
equal probability, i.e. at a rate of $\rho_*/3$ each. Notice that by this
definition, point sources are ``sinks" as well, in that they also absorb
particles.

\heading{Profile prediction.}  Unlike boundary-driven systems, simply
letting domain size tend to infinity does not lead to a meaningful limit.  To
see this, consider a single point source of rate $\rho$ in the center of
a roughly circular array of cells.  In the continuum limit, the jump
rate $J$ satisfies the Laplace equation with the condition $J=0$ on the
boundary of the domain and $J=\rho$ at the
location of the point source.  The only solution is $J\equiv 0$ almost
everywhere.  Similar considerations show that quantities like mean
particle count are also meaningless in this limit.

Nevertheless, one can still apply the prediction scheme for finite-size
systems: we simply solve
the discrete $J$ and $Q$ balance equations without taking continuum
limits.\footnote{Another possibility is to obtain a nontrivial continuum
  limit by allowing the injection rate $\rho$ to scale with domain size.
  We do not consider this possibility further.}  These equations take
the form
\begin{displaymath}
J_c = \frac13 \sum_{c'\sim c} J_{c'}~ \quad {\rm and} \quad Q_c = \frac13 \sum_{c'\sim
  c} Q_{c'}~,
\end{displaymath}
where $\sim$ is the neighbor relation on the hexagonal grid and the
equation holds for all cells $c$ that do not contain a point source.  As
before, open boundaries are implemented by imagining a strip of cells
$c$ just outside the array and setting $J_c=Q_c=0$ for these cells.
Finally, a cell $c$ containing a point source is implemented by {\em
  defining} the values of $J_c$ and $Q_c$ for that cell by
\begin{displaymath}
  J_c = \rho_*~\quad {\rm and} \quad Q_c = \frac32 \rho_*~T_*~,
\end{displaymath}
where $(T_*,\rho_*)$ are the parameters for the point source in cell
$c$~.

With the exception of $\kappa$, which depends explicitly on the geometry
of cells, all profiles can be derived from $J$ and $Q$ in exactly the
same way as before.  The expression for $\kappa$ here is easily found to
be
$$ \kappa(x,y) =2\sqrt\pi\cdot \frac{\sqrt3/4-\pi(\frac12 R^2 +
  r^2)}{1-2R} \cdot \frac{J(x,y)}{3\sqrt{T(x,y)}}
$$

\begin{figure}

  \begin{small}
    \begin{center}
      \begin{tabular}{cc}
        \includegraphics*[scale=0.8,bb=0.2in 0in 4.4in 2.8in]{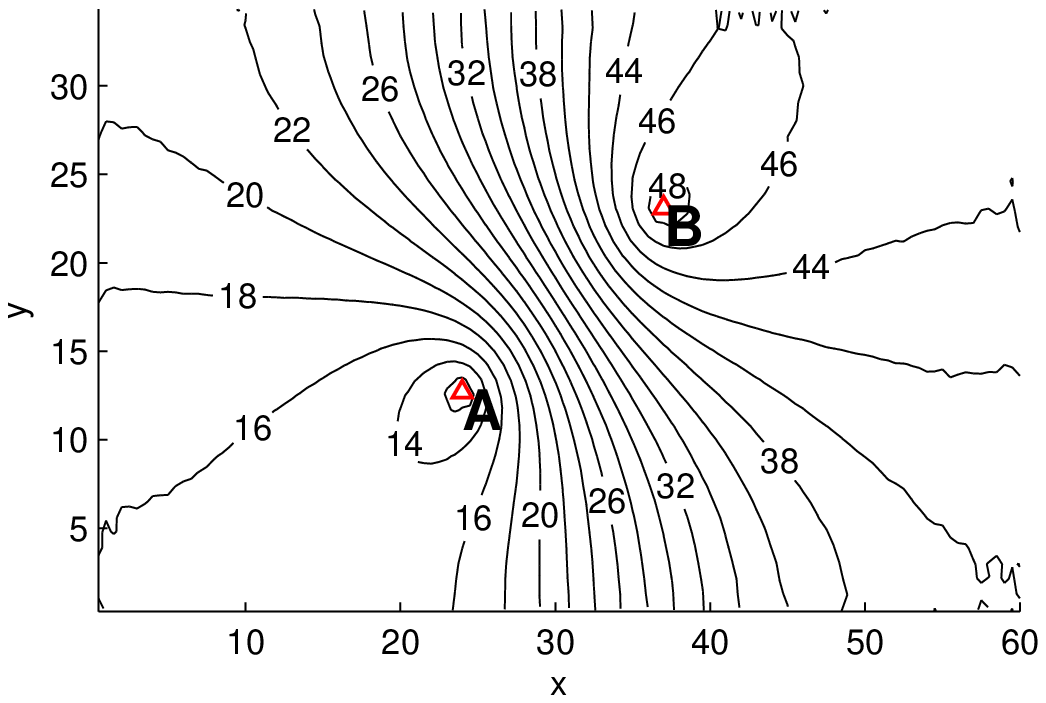}&
          \includegraphics*[scale=0.8,bb=44 0 313 203]{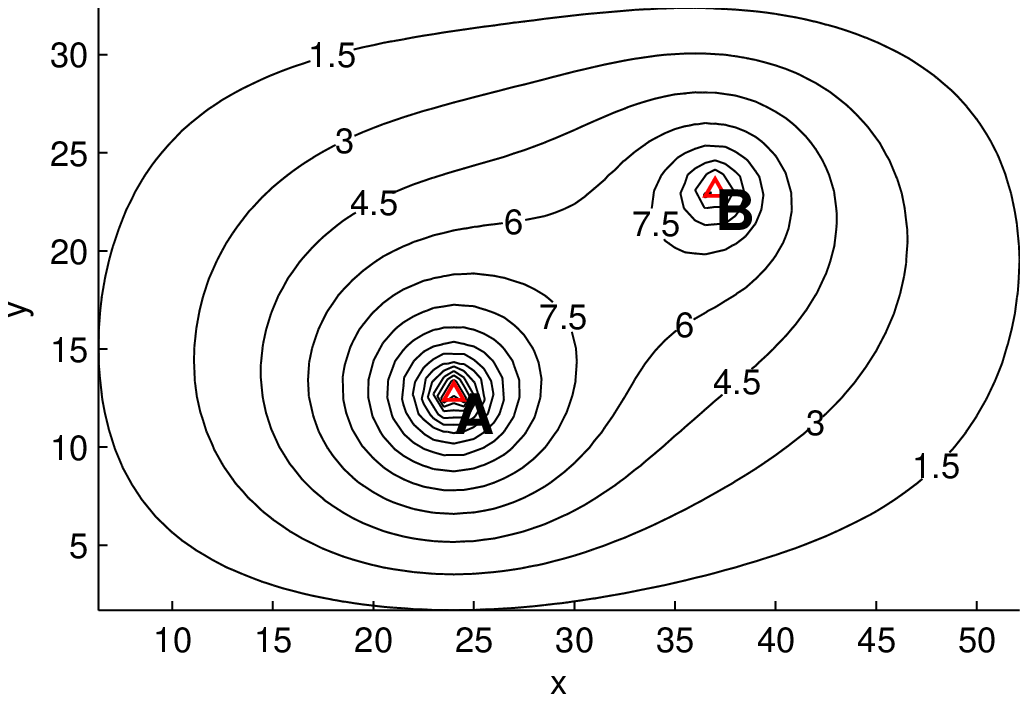}\\
        (a) Mean disk energy $D(x,y)$&
        (b) Mean particle count $\kappa(x,y)$\\[2ex]
      \end{tabular}\\[1ex]
    \end{center}
  \end{small}
  
  \caption{Simulation results for the hexagonal lattice with two point
    sources.  Source $A$ has parameters $(T,\rho)=(20,6)$; Source $B$,
    $(T,\rho)=(100,70)$.  The domain consists of 60 cells along the top
    and bottom edges and 40 cells along the left and right.  (Because of
    the hexagonal geometry, this means $120\times 40 = 4800$ cells in
    total.)  The discrepancy away from boundaries are $\leq 1.8\%$ for
    $D$ and $\leq 3.5\%$ for $\kappa$. }

  \label{fig:6}
\end{figure}

\heading{Simulation results.}  Fig.~\ref{fig:6} shows the results of a
simulation.  The domain consists of a hexagonal array with an
approximately rectangular boundary, with $60$ cells along the bottom
edge and $40$ along the left.  The fixed disks have radii $R=0.485$, the
rotating disks $r=0.05$, so that each particle hits the rotating disk on
average about 10 times before exiting.  We place two point sources in
the system at the indicated locations.  Shown are the disk energy
profile $D(x,y)$ and the mean particle count $\kappa(x,y)$.  The
discrepancy between predicted and simulation profiles are $\leq 1.8\%$
for $D(x,y)$ and $\leq 3.5\%$ for $\kappa(x,y)$ away from
boundaries. %

These results are as expected: the disk energy and particle count
profiles clearly reflect parameters of the point sources.  One also sees
clearly that the number of particles tends to zero as one approaches the
boundary of the domain, as it should for open boundaries.  The sparsity
of particles explains the relative noisiness near the boundaries of
Fig.~\ref{fig:6}(a) (as discussed in relation to the left edge in
Example 2).

\subsection*{Concluding remarks for Section 2}

We have extended the recipe for profile prediction proposed in \cite{ey}
to a variety of situations and tested it in many concrete examples.  In
each of the three examples presented (and the many others not shown), we
have found that (i) the predicted profiles are very simple to compute,
and (ii) the discrepancies between predicted and simulated values are no
more than a few percentage points.  Given that our parameter choices
take the systems tested very far out of equilibrium and the system sizes
considered are only moderate, we find the agreement between predicted
and simulated results strong.  We conclude that the scheme proposed is
effective for the class of models studied.


\section{Memory, finite-size effects, and geometry}
\label{sect:memory}

The success of the prediction scheme proposed in \cite{ey} and extended
in Sect.~2 of the present paper prompts the following question: Why is
it that memory and finite-size effects have so little impact on
macroscopic profiles in the models considered?  The purpose of this
section is to investigate, in the simpler setting of 1D chains, which
dynamical mechanisms and/or aspects of cell geometry are responsible for
this outcome, and what happens if the geometry is varied.

\smallskip

We focus on the following two issues:
\begin{enumerate}

\item {\em Directional bias due to memory of particle history,} i.e.,
  depending on its past history and the state of the cell, a particle
  may have a preferred direction for exiting; and

\item {\em Incomplete equilibration within cells,} which may (or may
  not) happen when particles collide too few times with the rotating
  disk before exiting.

\end{enumerate}

\subsection{Directional bias due to particle history}
\label{sect:3.1}

The first part of this subsection has some overlap with \cite{eckmann},
which contains a phenomenological study of the same bias questions.  Our
findings are consistent with those in that earlier study, though we take
a slightly different approach, and the bias-correction equations we derive
(which are simpler than those in \cite{eckmann}) will be tested in
models with significant bias.

The 1D-chain used here is exactly one row of the 2D-rectangular
configuration used in the previous section, with the top and bottom gaps
of each cell sealed, i.e., particles reflect off these two straight
segments elastically.  In Parts (A) and (B) below, we consider systems
with fixed disk radius $R=0.95$ (leading to gap size $=0.1$) and
rotating disk radius $r=0.2$, the same parameters as those used in
Examples 1 and 2 of the previous section.  These parameters will be
varied, with consequences, in Part (C).

The analysis below applies only to situations when there is a sufficient
number of particles in the cell. We assume a mean occupation number
of at least about 5.

\subsubsection*{(A) Single cell}

We begin by measuring the amount of bias in a single cell using
numerical simulations.  Consider a $2\times 2$ array of fixed disks
defining a single cell as shown. Heat baths with the following
parameters are attached to the left and right openings:
\begin{center}
  \begin{small}
    \begin{tabular}{ccc}
      &\includegraphics[bb=20 108 271 147,scale=0.6]{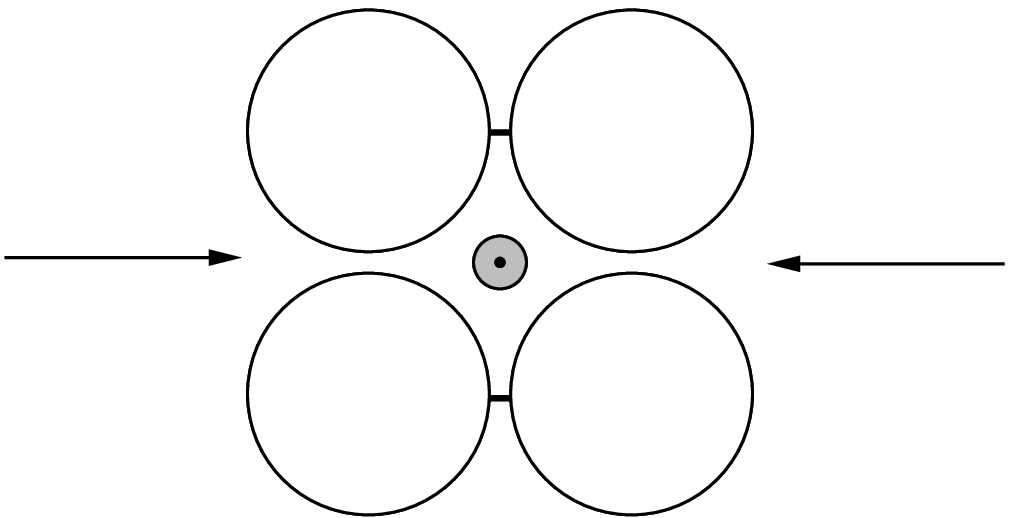}&\\
      $\tl=1-\frac12\dt^{in}$&&$\tr=1+\frac12\dt^{in}$\\[1ex]
      $\rl=1-\frac12\dr^{in}$&&$\rr=1+\frac12\dr^{in}$\\
    \end{tabular}
  \end{small}
\end{center}
\vspace*{30pt}

\noindent Let $\rho^-$ and $\rho^+$ denote the rates at which particles
exit the cell to the left and to the right respectively.  Similarly, we
denote mean energy outflux rates by $Q^\pm$, and define the mean
``temperatures'' of outgoing particles by $T^\pm = \frac23 Q^\pm /
\rho^\pm$~.  We express the observed bias by
$$
\dt^{out} = T^+ - T^-~,\qquad 
\dr^{out} = \rho^+ - \rho^-~.
$$
Note that zero directional bias would imply $\dr^{out}=0$~.


%

We then express $\dt^{out}$ and $\dr^{out}$ in terms of
$\dt^{in}$ and $\dr^{in}$, stipulating that $T_L, T_R$, and also
$\rho_L, \rho_R$, be symmetric about $1$. 
Expanding to leading order, we obtain
\begin{equation}
  \begin{array}{rlc}
      \dt^{out} &\approx& \at\dt^{in} + \bt\dr^{in}~,\\[1ex]
    \dr^{out} &\approx& \ar\dt^{in} + \br\dr^{in}~,
  \end{array}
  \label{eq:linear response}
\end{equation}
where the coefficients are computed (using $-0.2\leq \dt^{in}, \dr^{in}
\leq 0.2$) to be\footnote{The error estimates given here are $2\times$
  standard error (corresponding to 95\% confidence intervals).  This
  degree of uncertainty has almost no perceptible effect on the bias
  correction results in Sect.~3.2(B).}  
\begin{align*}
  &\at\approx 0.130\pm 0.001~,~~~ \bt\approx 0.014\pm 0.008~, \\[1ex]
  &\ar\approx 0.021\pm 0.007~,~~~ \br\approx 0.125\pm 0.001~.
\end{align*}

%

\heading{Relevant dynamical mechanisms.}  In general, we think of the
rotating disk, whose energy fluctuates wildly, as a randomizing force,
causing particles to lose memory of their previous spatial locations and
directions.  Without a doubt, this mechanism is responsible for
diminishing bias in exit direction.\footnote{Concave cell walls also
  help in the scattering of particles, but they appear not to be as
  important as the randomizing effects of the rotating disk.}  There is
a situation, however, when this mechanism fails: Suppose a particle
enters from the right and hits the rotating disk at roughly a right
angle. If the rotating disk at that moment happens to have energy
considerably lower than that of the particle, then after the energy
exchange (see Sect.~1.1), the normal component of the velocity of the
particle continues to dominate its tangential component, and the
particle is reflected back essentially the way it came.\footnote{A
  similar observation is made in \cite{eckmann}.}

Define a {\it reflected particle} to be one which, in the course of its
journey through the cell, hits only the rotating disk and the two fixed
disks on the side of the opening through which it enters, i.e. it is
``reflected back". (Such particles need not exit the first time around;
they can turn around and have multiple interactions with the rotating
disk.) For $\dt^{in}\equiv 0$ and $\dr^{in}\in[-0.2,0.2]$~, we find that
among particles entering from the left (respectively, the right), a
fraction  $r \approx 0.123$ is reflected. 
Since higher energy particles are more likely to reflect, one naturally
expects  the mean temperature of reflected 
particles to be higher than
that of other particles. At $(T,\rho)=(1,1)$, we find this difference to be
$t \approx 0.09$.

Assuming that particles that are not reflected are thoroughly randomized, 
one would expect, in terms of orders of magnitude, that 
$\alpha_T~,~\beta_\rho \sim r$ and $\beta_T \sim rt$. 
The estimate for $\beta_T$ is  based on the fact
that on the side with higher injection rate, a larger fraction of particles is
reflected, {\it and} this fraction has higher temperature.
Computed values of 
$\alpha_T~,~\beta_\rho$ and $\beta_T$ are consistent with these
very na\"ive estimates.
As for $\alpha_\rho$,
we can only say that it should be $>0$ because higher temperature
leads to more reflected particles. 

%

The findings above support our contention that
reflected particles are an important source of bias, but the true story
is much more complicated. For example, when a particle interacts with
the rotating disk, even as some memory is lost, its post-collision trajectory
is not entirely random.
It also retains some
memory of its previous energy since the normal component of its velocity
upon impact is kept. A complete analysis is beyond the scope of this
article.
%




\heading{Linear response for general $(T,\rho)$.} If bias arises mainly
from individual particle trajectories --- and that appears to be the
case when there is a sufficient number of particles in the cell ---
$\dr_{out}$ and $\dt_{out}$ should scale linearly with the total
injection rate $\rl+\rr$.  This allows us to extend the linear response
relation~(\ref{eq:linear response}) to boundary conditions
$$
(\tl,\rl)=(T-\frac12\dt^{in},\rho-\frac12\dr^{in})~,\qquad
(\tr,\rr)=(T+\frac12\dt^{in},\rho+\frac12\dr^{in})
$$
for general $(T,\rho)$. The general linear response relations
are
\begin{align}
  \frac{\dt^{out}}{T} &\approx \at\cdot\frac{\dt^{in}}{T} + \bt\cdot\frac{\dr^{in}}{\rho}~,\nonumber\\[2ex]
\frac{\dr^{out}}{\rho} &\approx \ar\cdot\frac{\dt^{in}}{T} + \br\cdot\frac{\dr^{in}}{\rho}~,
  \label{eq:general linear response}
\end{align}
where the phenomenological coefficients $\at$, $\bt$, etc., are as in
Eq.~(\ref{eq:linear response}).  We have numerically verified that these
scaling relations are accurate to within 2\% for $2\leq\rho\leq 12$ and
$10\leq T\leq 60$.\footnote{As pointed out in \cite{eckmann}, in
  principle these coefficients may depend on particle density (this
  follows from dimensional analysis).  We have not found evidence for
  any significant variation within the range of parameters used here.}

\subsubsection*{(B) Chains}

A natural next step is to try to use the linear response
relation~(\ref{eq:general linear response}) to account for the effects
of bias on macroscopic profiles.  {\it A priori}, knowledge of bias effects 
in a single cell need not tell us about how a cell would behave when 
placed within a chain: single cells are driven by
I.I.D. bath injections whereas cells within a chain receive correlated
inputs, and long-range correlations are known to be substantial
in nonequilibrium steady states.

\heading{Prediction scheme with built-in bias correction.}  We derive
phenomenological equations for macroscopic profiles {\em assuming} 
that bias at a given site $i$ is determined solely by the temperature, 
particle flux, and their gradients at $i$, in the same way as in a single cell.  Let $Q_i^\pm$ and $J_i^\pm$ be the mean rates of energy and particle flux
out of cell $i$, i.e., $Q^+_i$ is the mean total energy that passes
from cell $i$ to cell $i+1$ per unit time, and so on, 
and let $T_i^\pm = \frac23 Q_i^\pm / J_i^\pm$~.  

First,
we have the balance equations
\begin{align}
  \rho_{i}^+ + \rho_{i}^- &= \rho_{i-1}^+ + \rho_{i+1}^-~, \nonumber\\[0.5ex]
  Q_{i}^+ + Q_{i}^- &= Q_{i-1}^+ + Q_{i+1}^-~.
  \label{eq:bias-corrected1}
\end{align}
These are augmented by bias corrections of the same form as
Eq.~(\ref{eq:general linear response}):
\begin{align}
  \frac{T_{i}^+ - T_{i}^-}{T_i} &= \at\cdot\frac{\dt_{i}^{in}}{T_i} + \bt\cdot\frac{\dr_{i}^{in}}{\rho_i}~,\nonumber\\[2ex]
  \frac{\rho_{i}^+ - \rho_{i}^-}{\rho_i} &= \ar\cdot\frac{\dt_{i}^{in}}{T_i} + \br\cdot\frac{\dr_{i}^{in}}{\rho_i}~,
      \label{eq:bias-corrected2}
\end{align}
where we define
$$
T_i =\frac12(T_{i-1}^++T_{i+1}^-)~,\qquad 
\rho_i = \frac12(\rho_{i-1}^++\rho_{i+1}^-)$$
and
$$
  \dt_{i}^{in} = T_{i+1}^--T_{i-1}^+~,\qquad
  \dr_{i}^{in} = \rho_{i+1}^--\rho_{i-1}^+~.
$$
Our proposal is to first compute the coefficients $\at$, $\bt$, $\ar$, $\br$ from single cell simulations, and then to solve 
Eqs.~(\ref{eq:bias-corrected1}) and
(\ref{eq:bias-corrected2}) numerically to obtain bias-corrected
predictions.  If
$\ar=\br=\at=\bt=0$, these equations are precisely the prediction
scheme outlined in Sect.~\ref{sect:1.2}.

We have carried out simulations to validate this bias correction
scheme. For chains with the cell geometry in Part (A), the effect of
bias is generally small but measurable.  Even with these more delicate
measurements, our scheme gives quite accurate predictions of
$\dr^{out}(x)$ and $\dt^{out}(x)$~.  These findings will be corroborated
by those in simulation studies for which bias is more significant, which
we now discuss.

\subsubsection*{(C) ``Good'' geometry, ``bad'' geometry}

\begin{figure}
  \begin{small}
    \begin{center}
      \begin{tabular}{cp{2ex}cp{2ex}c}
        \includegraphics*[scale=0.5,bb=0.95in 0.93in 2.95in 2.94in]{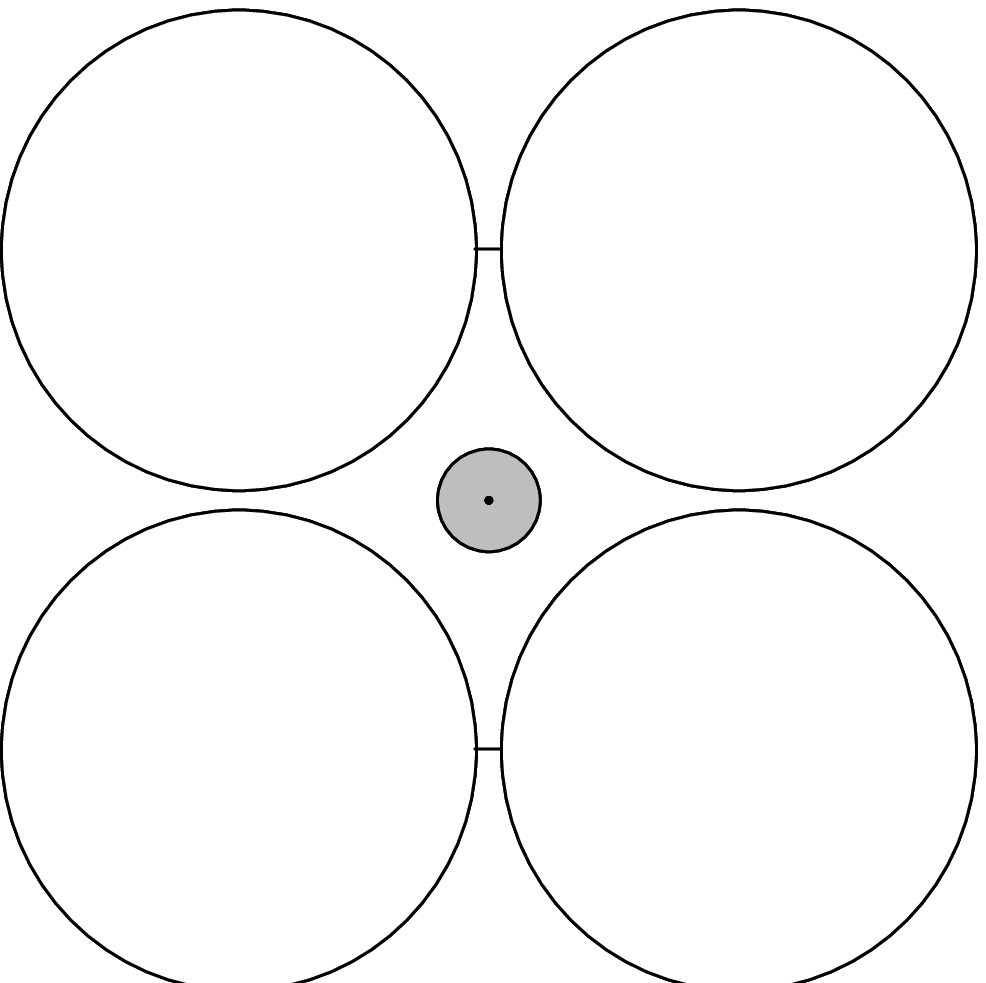} &&
        \includegraphics*[scale=0.5,bb=0.95in 0.93in 2.95in 2.94in]{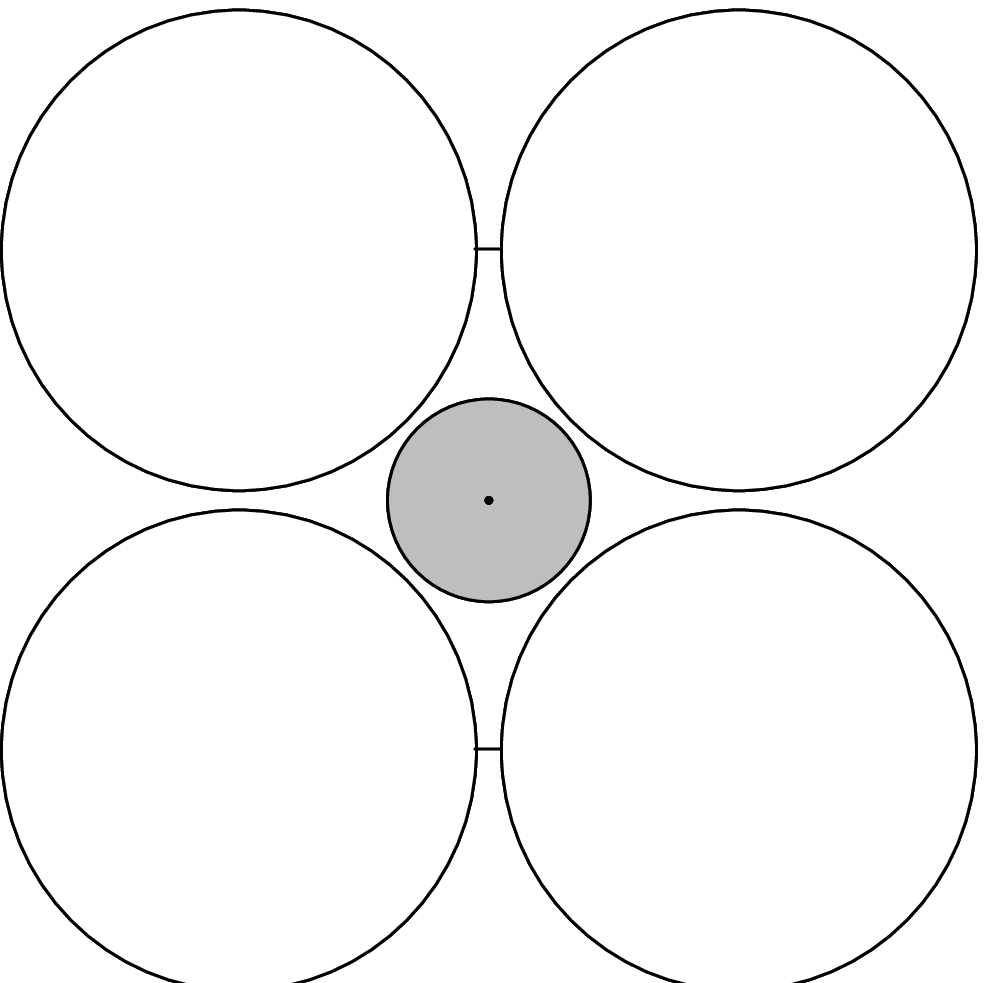} &&
        \includegraphics*[scale=0.5,bb=0.5in 0.48in 2.5in 2.49in]{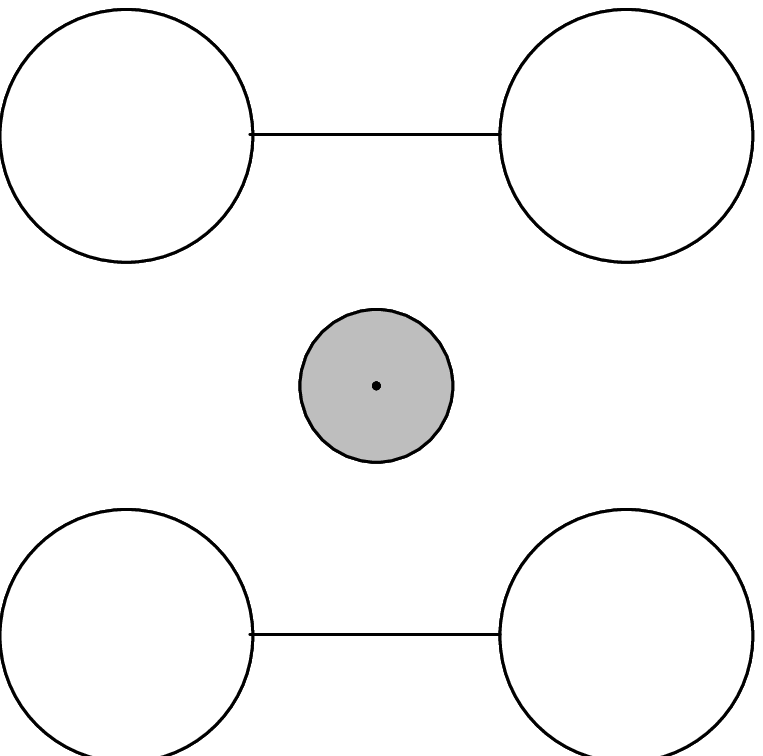} \\[1ex]
        (a) $R=0.95~,~~r=0.2$ &&
        (b) $R=0.95~,~~r=0.4$ &&
        (c) $R=0.5~,~~r=0.3$ \\[1ex]
      \end{tabular}
    \end{center}
  \end{small}
  \caption{Examples of different cell geometries.  Panel (a) shows the
    geometry used in most examples in this paper, including Examples
    1--3.  The geometry in (b) leads to strong directional bias.  In
    (c), we show a geometry that is likely to lead to incomplete
    equilibration between particles and the rotating disk.}

  \label{fig:8}
\end{figure}

Up until now, we have considered a fixed cell geometry (namely that in
Fig.~\ref{fig:8}(a)), which by all counts gives rise to very little
bias. One would expect bias to increase for geometries which have a
larger fraction of reflected orbits. For example, one may guess that if
one increases the gap size $|\gamma| = 2(1-R)$, it will be easier for
reflected particles to exit, and the bias coefficients will go up.  We
have found this to be true, provided $|\gamma|$ is somewhat smaller than
$2r$.

A more straightforward way to increase the fraction of reflected
particles, however, is to fix gap size and increase $r$, for larger
rotating disks are more effective in {\em blocking} incoming orbits.
Table \ref{tab:1} shows the values of $\at$ and $\bt$ for various values
of $r$ with gap size fixed at $0.1$, equivalently with $R=0.95$. The
trends are self-evident.

\begin{table}
  \begin{center}
    \begin{small}
      \begin{tabular}{c|ccccc}
        r & 0.055 & 0.1 & 0.2 & 0.3 & 0.4 \\\hline
        $\at$ & $7\times 10^{-4}$ & 0.060 & 0.13 & 0.23 & 0.33 \\
        $\bt$ & $1.4\times 10^{-3}$ & 0.004 & 0.014 & 0.023 & 0.028 \\
      \end{tabular}
    \end{small}
  \end{center}
  \caption{Single-cell bias coefficients vs. rotating disk radius.  The
    gap size is $|\gamma|=0.1$.  We observe similar trends for $\ar$ and
    $\br$.  Below $r_{\min}=0.05$,
particles can fly through a cell without being in contact with any wall
or disk (infinite horizon), and at $r_{\max}\approx 0.46$, the rotating disk touches the fixed disks.} 
  \label{tab:1}
\end{table}

\begin{figure}
  \begin{center}
    \begin{tabular}{cp{6ex}c}
      \includegraphics[bb=0in 0in 2.25in 2.25in]{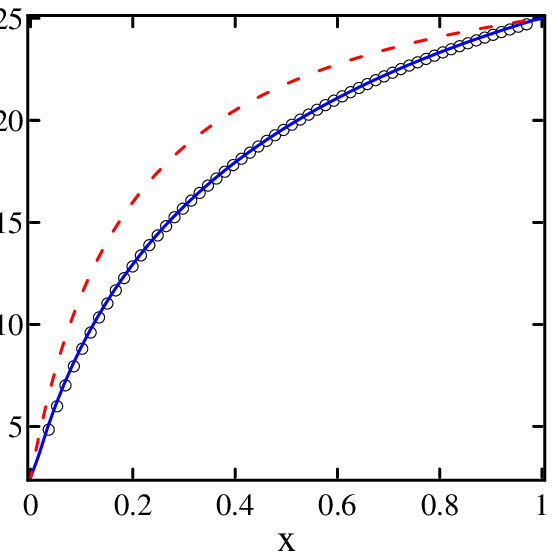} &&
      \includegraphics[bb=0in 0in 2.25in 2.25in]{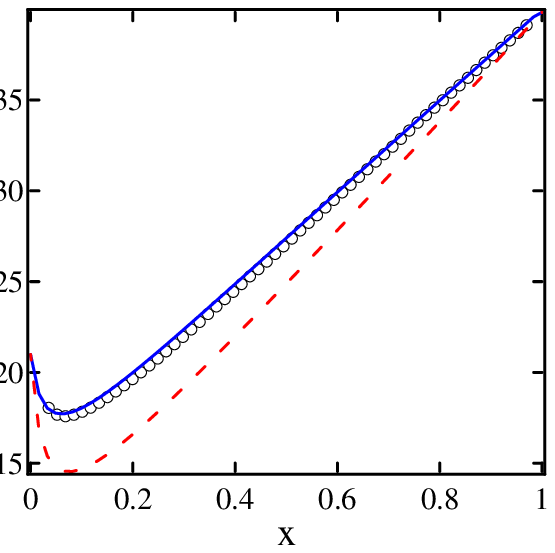} \\
      (a) Mean disk energy $D(x)$ &&
      (b) Mean particle count $\kappa(x)$ \\
    \end{tabular}
  \end{center}
  
  \caption{Profiles for chains with ``bad'' geometry.  Here, the chains
    have rotating disk radius $r=0.4$ (we keep $R=0.95$).  Shown are
    simulation data (circles), predicted profile without bias (dashed
    lines), and bias-corrected predictions (solid lines).}
  \label{fig:9}
\end{figure}

\heading{Example 4: A chain with ``bad" geometry.}  We consider here
cells with $R=0.95$ and $r=0.4$; a picture of such a cell is shown in
Fig.~\ref{fig:8}(b). This is a candidate for a ``bad" geometry: it is
almost as easy for a particle to bounce back and forth between two
adjacent cells as it is to pass from a side chamber to the top or bottom
chamber of the same cell. Simulations were performed for a chain with
$N=60$, $\tl=5$, $\tr=50$, $\rl=2$, and $\rr=12$; the results are
displayed in Fig.~\ref{fig:9}.

Predicted profiles using the scheme {\em without} bias correction are
shown in dashed lines, while bias-corrected profiles are shown as solid
lines. As can be seen from the plots, deviations of simulation data
(circles) from uncorrected profile predictions are nontrivial, unlike
the situations in Examples 1--3, where cell geometries are considerably
more favorable. On the other hand, the fit with profiles from our
bias-corrected prediction scheme is excellent.  These profiles are
computed from Eqs.~(\ref{eq:bias-corrected1}) and
(\ref{eq:bias-corrected2}) using coefficients $\at$, $\bt$, $\ar$, and
$\br$ from single cell simulations with $r=0.4$.

\subsection{Incomplete equilibration}
\label{sect:3.2}

When every particle spends a very long time in a cell before moving on,
one may expect the particles and the rotating disk in the cell to come
close to equilibrating.  Thus one may think that small gaps (or
passageways) between cells speed up the convergence to local
equilibrium.  In \cite{ey} and \cite{eckmann}, these gaps were taken to
be very small: on average, a particle hits the rotating disk more than
$12$ times on each visit to a cell.  In Examples 1--3 in the present
paper, these conditions are relaxed to 3-5 hits per visit.  But are
small gaps a necessary condition?  Put differently, suppose the gaps are
large (as in Fig.~\ref{fig:8}(c)), so that each particle moves through
every cell quickly, hitting the rotating disk only once or twice (or
even less) on each pass.  Would such incomplete equilibration
necessarily lead to anomalous behavior?

We now attempt to address this question.  The short answer is: A
particle's failure to completely equilibrate with its cell environment
on each pass is, in and of itself, {\it not} an obstruction to normal
behavior.  However, large gaps can amplify {\em finite-size effects}.

For lack of a better term, let us call a cell boundary ``porous'' if a
particle passing through the cell is not likely to collide with the
rotating disk too many times before exiting. For the type of models
discussed earlier, porosity is equivalent to larger gap size.  We
propose below a connection between porosity and finite-size effects, and
provide numerical confirmation of our thinking.

\subsubsection*{Dynamics of equilibration and finite-size effects: a heuristic discussion}

From the point of view of a particle, equilibration takes place as
follows: It moves about the chain in what is essentially a random walk,
exchanging energy with the rotating disks with which it collides, hence
indirectly with other particles which collide with the same disks.  It
is an ongoing process, a very messy one involving large fluctuations.
To say that a particle has ``equilibrated'' basically means that it has
acquired ``the right statistics.''  From this picture, it is clear that
full equilibration in one cell before proceeding to the next is not at
all necessary.

However, a particle has to remain in the system long enough to fully
sample the totality of its surroundings. For a concrete example, suppose
we place a thermostat in the middle of a chain.  As noted in Example 2,
particles retain their normal components in collisions, one might
naively expect that a particle has to interact with the thermostat many
times for equilibration to occur.  (Actually, it can also acquire this
information indirectly.)  If the chain is too short and most of the
particles exit too quickly, then the thermostat may not be fully
effective due to insufficient sampling.
In the case of boundary injections, if the domain is not large enough
for particles to collide a sufficient number of times before exiting the
system, then the different sets of statistics carried by injected
particles (e.g., bath temperature) may not have enough opportunity to
get absorbed and propagated through the chain.

The phenomena described above can be thought of as {\it incomplete
  equilibration with thermostats or baths}.  They will lead to
finite-size effects. This kind of equilibration does not require that
the particle make many collisions in each cell that it visits, though
more collisions are conducive to better statistics.

Here is how porosity enters. To exaggerate the problem, consider a
situation where cell boundaries are sufficiently porous that on average
particles travel ballistically for some (microscopic) distance without
exchanging energy with any disks. In such a system, particles will need
more ``room'' to make the same number of collisions. The more porous the
cell boundaries, the greater the mean free path. The discussion above
continues to apply, but a longer chain or larger domain is required to
ensure complete equilibration with thermostats or baths.

Alternatively, consider a chain of length $N$ in which the mean free
path of particles is $\ell$.  If we view particles in this chain as
moving on the interval $(0,1)$, the paths traced out will resemble those
of a diffusion whose associated lengthscale increases with $\ell$ (for
fixed $N$) and decreases with $N$ (for fixed $\ell$). This suggests the
following: (a) Incomplete equilibration in an $N$-chain can have a
smoothing effect on the profiles of relevant observables on $(0,1)$, as
though they are convolved with some kernel.  In particular, profiles
that are strongly curved may lose some of their curvature.  (b) Whatever
the porosity, this smoothing vanishes as $N\to\infty$.

The discussion above is entirely heuristic, but we believe it captures
the flavor of part of what is going on.  To summarize, our reasoning
implies the following: (1) Other things being equal, the more porous the
cell geometry, the larger the deviation from expected behavior for a
fixed system size.  (2) In the infinite-volume limit, these effects will
vanish, i.e., these are genuinely {\em finite-size} effects.

\bigskip
A number of numerical studies were carried out to support (or refute) the thinking above. Chains driven by (unequal) boundary injections
behave exactly as described: finite-size effects in the form of decreased 
curvature are clearly visible but 
for the most part not too pronounced for chains of moderate length. 
The thermostat example 
discussed above provides a more dramatic illustration of complete {\it versus}
incomplete equilibration. 

\begin{figure}
  \begin{small}
    \begin{center}
      \includegraphics[scale=0.6,bb=0in 0in 6.0in 0.6666666666666666in]{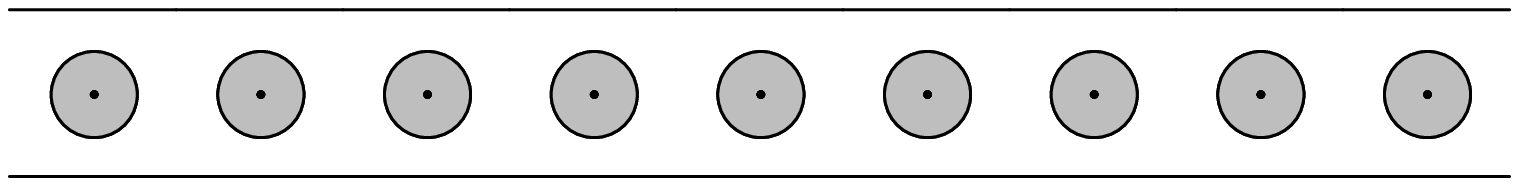}\\
      (a) Cells without borders\\[5ex]
      \hspace*{-5ex}
      \begin{tabular}{cp{3ex}c}
        \begin{tabular}{cc}
          
          $N=31$ & $N=401$ \\[2ex]

          \resizebox{!}{1.5in}{\includegraphics*[bb=0in 0in 2.5in 2.25in]{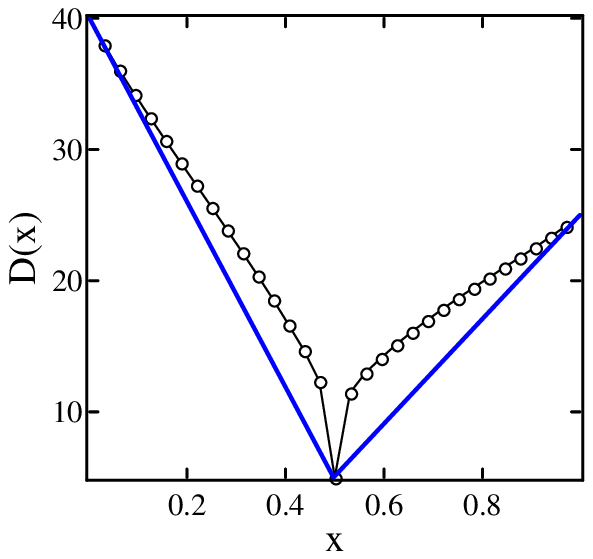}} &
          \resizebox{!}{1.5in}{\includegraphics*[bb=0in 0in 2.5in 2.25in]{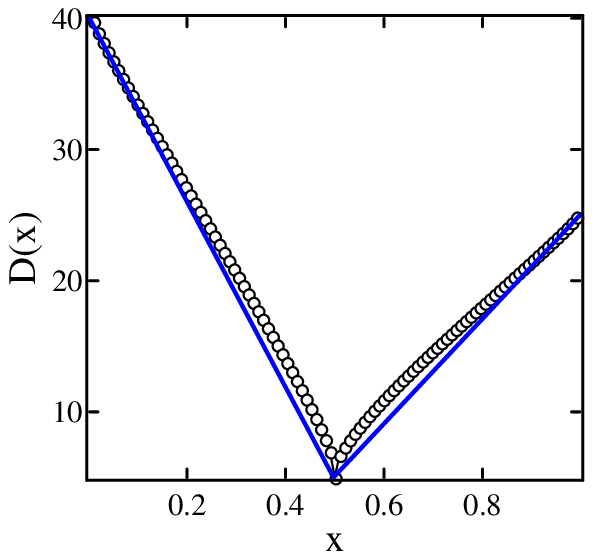}} \\[2ex]

          \resizebox{!}{1.5in}{\includegraphics[bb=0in 0in 2.5in 2.25in]{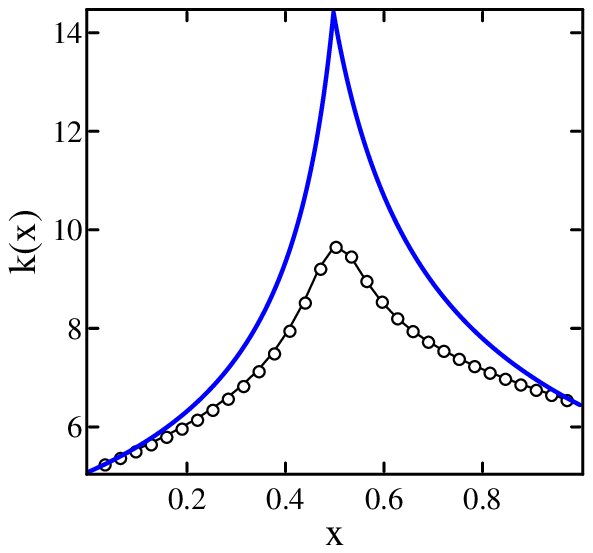}} &
          \resizebox{!}{1.5in}{\includegraphics[bb=0in 0in 2.5in 2.25in]{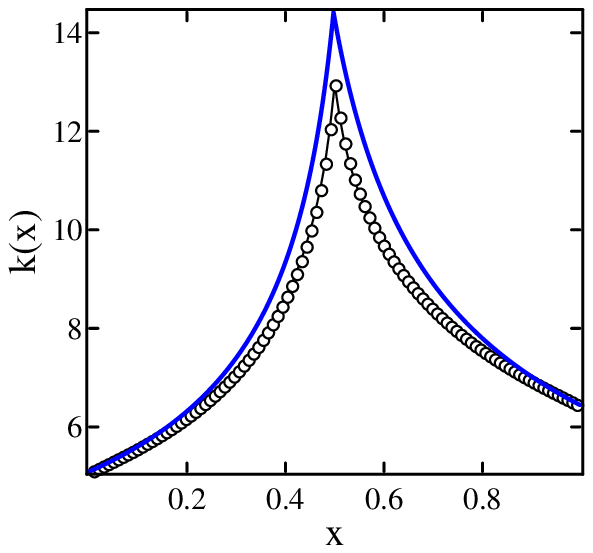}} \\

        \end{tabular} &&

        \includegraphics[scale=0.9,bb=0in 1.5in 2.25in 3in]{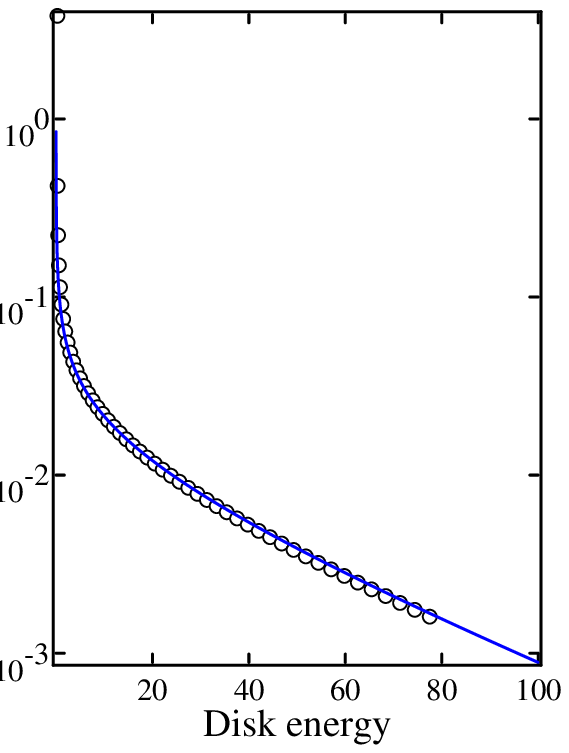} \\
        
        (b) Profiles && (c) Disk energy distribution, $N=31$ \\
      \end{tabular}\\[3ex]
    \end{center}
  \end{small}

  \caption{A 1D chain without cell borders.  A thermostat with
    temperature $T_*=10$ is placed at exactly $x=0.5$; the boundary
    conditions are $\tl=80$, $\tr=50$, $\rl=\rr=8$; $r=0.5$.  Panel (a)
    illustrates the geometry.  In (b), we plot mean disk energy and
    particle count profiles for two values of $N$, showing in each case
    both simulation data (curves with open circles) and predictions
    (solid lines).  Note that the value of the disk energy profile
    $D(x)$ at $x=1/2$ is fixed by the thermostat to $\frac12 T_* = 5$.
    Panel (c) shows the energy distribution for the 10th rotating disk
    from the left for $N=31$; empirical data (open circles) are plotted
    against the disk energy density for $\mu^{T,\rho}$ with $\rho=8$ and
    $T\approx 44.3$, the empirical temperature of the disk.}
  \label{fig:10}
\end{figure}

\heading{Example 5: Cells without borders.}  The basic setup is
illustrated in Fig.~\ref{fig:10}(a): it consists of a 1D chain connected
to two heat baths at the two ends and with a single thermostated cell in
(exactly) the middle.  The boundary conditions are $\tl=80$, $\tr=50$,
and $\rl=\rr=8$, and the thermostat is set to temperature $T_*=10$.
Neglecting the possible effects of bias, the predictions in
Sect.~\ref{sect2} give, in the continuum limit, a piecewise linear
profile for the disk energy $D(x)$: it linearly interpolates the 3
values $D(0)=40$, $D(\frac12)=5$, and $D(1)=25$; particle count
$\kappa(x)$ is computed as before, i.e. assuming the system is
partitioned into cells (with invisible separations), with each cell
containing exactly one disk.  System size will be varied in this
study. We consider the following three cell geometries:

\begin{itemize}

\item {\em Geometry A (small gaps):} $R=0.95$, resulting in gap size
  $|\gamma|=0.1$, and $r=0.2$; see Fig.~\ref{fig:8}(a).  These are the
  parameters used in Examples 1--3. The average number of interactions
  with the rotating disk is $\approx 6$ per visit.

\item {\em Geometry B (large gaps):} $R=0.5$, resulting in gap size
  $|\gamma|=1$, and $r=0.3$, as shown in Fig.~\ref{fig:8}(c). This
  configuration has infinite horizon. Number of hits per pass is
  $\approx 0.9$.

\item {\em Geometry C (no walls between cells):} $R=0$ and $r=0.5$.  A
  chain with this geometry is shown in Fig.~\ref{fig:10}(a). This is as
  ``porous" a geometry as there can be.

\end{itemize}

For each of these geometries, we computed $D(x)$ and $\kappa(x)$ for
$N=$31, 61, 121, 201 and 401 (for Geometry C, note that the derivations
in Sect.~\ref{sect:1.2} are not affected by the fact that there are no
cell walls).  In each case, deviation from the predicted profile
decreases monotonically as the length of the chain $N$ increases.  Two
sets of results for Geometry C are shown in Fig.~\ref{fig:10}(b).  As
one can see, finite-size effects are very nontrivial at $N=31$, and are
still clearly visible at $N=401$.  From the plotted values of $D(x)$
(set to $5$ at $x = \frac12$ on account of the thermostat), one sees
that the effectiveness of the thermostat is seriously compromised at
$N=31$: $D(x)$ being higher than predicted is consistent with
$\kappa(x)$ being lower than predicted.  The plot of $\kappa$ shows
unambiguously the effects of smoothing, confirming the
incomplete-equilibration-with-thermostat picture proposed.  By contrast,
the plots for Geometry A (not shown) show no visible finite-size effects
beyond $N=61$; at $N=31$, the deviations are already smaller than those
for Geometry C at $N=401$.  Simulation results for Geometry B are
between those of A and C; they contain no surprises.

An intriguing question is the distribution of disk energy for this type
of chains, for they are the antithesis of chains with tiny gaps
(introduced in \cite{ey} with the idea that equilibration within cells
may lead to more rapid convergence to local equilibrium).  A specific
question is whether as $N \to \infty$ single site marginals of steady
states will tend to measures of the form $\mu^{T,\rho}$ defined in
Sect.~1.1. Numerical evidence suggests that they will: {\it We found
  that disk energy distributions are of Gibbs form for even fairly short
  chains}, independently of the completeness of equilibration with baths
or thermostats.  Fig.~\ref{fig:10}(c) shows the distributions of disk
energy at $x\approx 0.3$ for Geometry C with $N=31$.  The plot is in
log-linear scale; data collected from simulations are plotted against
disk energy distribution for $\mu^{T,\rho}$ with $\rho=8$ and $T$ equal
to the empirical temperature at the site in question.  Notice the
remarkable agreement, demonstrating how close these marginals are to
some $\mu^{T,\rho}$ long before $T$ approaches its continuum value.

\subsection{Implications for general model designs}

Building on previous ideas, our results in Sects.~2,\ \ 3.1, and 3.2
point to a very rich class of out-of-equilibrium Hamiltonian models for
which we have a simple and effective scheme for predicting approximate
energy and particle density profiles. We recapitulate here some of the
ideas, and discuss further implications.

\medskip
Retaining the particle-disk interaction introduced in \cite{rateitschak-klages} 
and \cite{llm}, we have shown that the ideas first put forward in \cite{ey} 
are applicable to a very flexible collection of models: 

First, the results are not limited to 1D; simulations have been carried
out in 1D and 2D, and we have seen no reason why the type of
  prediction scheme studied in this paper cannot be applied in three and
  higher dimensions, provided the models have suitable interactions
  between moving particles and pinned-down objects. 
The type of lattices over which to place the rotating disks appears not
to be an issue either; we have illustrated it for rectangular and
hexagonal lattices. We have tested a number of different ways to
maintain the system out of equilibrium: be it boundary driven (we have
tested a variety of boundary conditions), or driven from the interior,
using point sources or rows of thermostats -- in all cases our
prediction scheme has produced easy-to-compute profiles that show strong
agreement with simulations.

To understand why memory
and finite-size effects are so mild in the models tested in \cite{ey} and in
Sect. 2, we have proposed an explanation
in terms of geometric designs: In these models, the cells have mostly 
concave (scattering) walls, are connected by narrow  passageways, 
and have rotating disks at good distances away from the walls. 
We have shown in Sect. 3.1, as have the authors of \cite{eckmann} earlier, that
such designs give rise to very low degrees of directional bias. We have also 
explained in Sect. 3.2 that the small passageways, which facilitate equilibration
within cells, also ensure the efficient absorption of bath and thermostat 
information, reducing finite-size effects. This explains to some degree the
rather surprising fact that in \cite{ey}, for example, the prediction scheme 
gives remarkably good results in chains with a mere 30 cells.

Going beyond the models discussed in the last paragraph, we have found
bias to be small even as we relax considerably the parameters there,
i.e., Assumption 1 in Sect.~1.1 applies to a quite large collection of
models.  However, there are equally natural geometries that give rise a
non-negligible amount of bias.  For these geometries, we have derived a
correction scheme which is relatively straightforward to implement, and
shown it to give accurate results (see Sect.~3.1).  In a different
direction, we have produced numerical evidence in support of our ideas
on porosity of cell boundaries and finite-size effects (Sect.~3.2).  If
these heuristic ideas could be further developed and validated, the
implications would be that the results in \cite{ey} and Sect. 2 of this
paper will remain valid for systems with large gap sizes (perhaps no
walls at all) -- provided we go to larger system sizes.

We finish with two examples: Example 6(A), along with Example 5,
  represents one of the greatest departures from the models studied in
  \cite{ey} for which our prediction scheme remains effective; Example
  6(B), a close cousin of Example 6(A), makes a connection to the models
  studied in \cite{llm}.

\begin{figure}
  \begin{center}
    \begin{small}
      \begin{tabular}{cp{8ex}c}
        \includegraphics[scale=0.7]{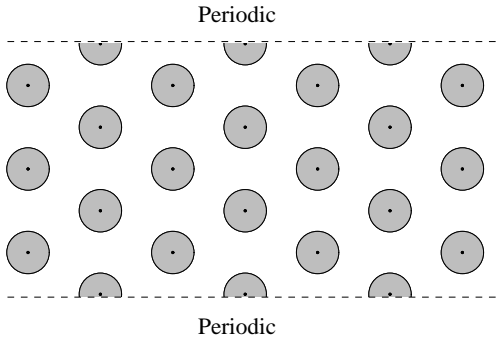}&&
        
        \includegraphics[scale=0.95,bb=0in 0in 2.5in 2.25in]{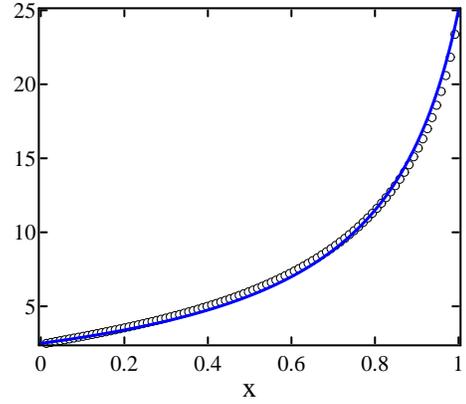}\\

        (a) Hexagonal grid without cell borders &&
        (b) Mean disc energies vs. $x$ \\
      \end{tabular}\\[2ex]
    \end{small}
  \end{center}
  \caption{Hexagonal model without cell borders.  Panel (a) illustrates
  (to scale) the geometry and placement of rotating disks. In (b), 
  we plot mean disk energies, showing 
    simulation data as a function of $x$ (open circles) and the
    predicted profile (solid line); the size of the array is 90 $\times$ 3.}
  \label{fig:10.5}
\end{figure}

\heading{Example 6.  (A)} We consider a cylindrical domain (or
rectangular domain with periodic boundaries in the vertical
direction). Inside this domain, rotating disks are placed on a hexagonal
lattice with no cell walls or partitions of any kind separating the
disks. The disks have radii $r=0.025$, and the centers of adjacent disks
are 1 unit apart. See Fig.~\ref{fig:10.5}(a).  Heat baths inject
particles into the system from the left and the right.  To deduce energy
and particle count profiles, we subdivide the domain into hexagonal
cells each enclosing exactly one rotating disk in its center, and
proceed as before.  Fig.~\ref{fig:10.5}(b) shows the computed disk
energy profile (open circles) and the predicted profile (without bias
correction) for a system consisting of a $90\times 3$ hexagonal array,
with $(\tl,\rl)=(5,12)$ and $(\tr,\rr)=(50,2)$.  A few other sets of
parameters besides this one were also tested, with results ranging from
good to very good.

\heading{(B)} The configuration in Fig.~\ref{fig:10.5}(a) is reminiscent
of the model used in \cite{llm}, except that the rotating disks in
\cite{llm} are placed much closer to each other. In (A), we have
deliberately kept them apart to avoid bias, which for this geometry can
be large: We have carried out a simple experiment similar to that in
Sect.~3.1(A) involving 3 rotating disks of radius $0.475$, separated by
3 narrow channels of width 0.05.  When all 3 disks are thermostated at
$T=1$, simulations show that a particle entering the triangular region
between the disks through one of the channels is 3-4 times more likely
to exit from the same channel than either one of the other two channels.
This suggests that a system such as the one in (A) but with disks packed
this close will exhibit significant bias.  Preliminary testing confirms
this thinking.  For example, when $\rl=\rr$, the profile for $J$, which
in the absence of bias would be constant along the length of the
cylinder, was found to be nontrivially curved (concave).

\heading{Remark.}  Finally, we observe that as far as methods and
overall ideas are concerned, no aspect of our studies is tied to the
particle-disk interaction used.  Thus it is reasonable to conjecture
that they may extend to other suitable short-range interactions.

\section{Systems with constant particle number}
\label{sect:closed systems}

Up to this point in the paper, we have studied
nonequilibrium steady states in systems where both particles and energy
can flow across the boundary, with bath injection rates $\rl$ and $\rr$
playing the role of chemical potentials.  In this section, we consider
systems in which the total particle count is maintained at a constant
number, and only energy is allowed to flow into and out of the system.
The difference between these particle-conserving chains and the chains
we studied earlier is akin to the difference between canonical and grand
canonical ensembles in equilibrium statistical mechanics.

To keep the number of particles fixed in models of the type studied in
\cite{ey} and in this paper, one way is to replace particles upon their
reaching the boundary of the domain with new particles whose energies
reflect those of the baths.  That is to say, each particle that exits is
replaced instantaneously by a new one, and there is no other infusion of
particles.  Another possibility is to isolate the system entirely from
its surroundings, and to place in different parts of the domain
thermostats set to unequal temperatures.
  Below, we report on findings using particle-conserving baths; we have
  found that simulations using thermostats yield similar
  results.

\subsection*{Example 7: 1D-chain with particles replaced on exit}

We consider a 1D $N$-chain as before, with cells $i=1$
and $i=N$ being in contact with heat baths. Particles are replaced {\em
  immediately upon exit.}  To be definite, we assume that when a particle 
exits the system at location $x \in \gamma$, where $\gamma$ is 
either the
  left opening of cell $1$ or the right opening of cell $N$, 
  independently of its exit velocity it is replaced by a new particle
  which appears in the same position $x$ and has a new velocity
$v'$ sampled from the bath distribution in Eq.~(\ref{eq:bath}).
Equivalently, one can think of the gaps leading outside the chain as
being sealed off by two ``thermostated walls,'' and when a particle hits
one of these walls, it is reflected but with a new velocity.  With these
particle-conserving dynamics, the total number of particles $M$ remains
constant in time.  Taking a continuum limit means letting $M,
N\to\infty$ while keeping the density $M/N$ constant.

\heading{Profile prediction.}  We submit that the various 
profiles can still be predicted following
the basic recipe in Sect.~\ref{sect:1.2}: one first solves for the mean
particle jump rate $J(x)$ and the mean energy outflux $Q(x)$; the other
profiles are then expressed in terms of $J$ and $Q$ as before.

What is different here is that because particles are replaced upon exit,
the boundary conditions for $J$ are $J_0=J_1$ and $J_{N}=J_{N+1}$~.
These (discrete) homogeneous Neumann boundary conditions lead, assuming
zero bias, to $J(x)\equiv J_0$, i.e., the $J$ profile is constant in
space.  The $Q$ profile has the same boundary conditions as before and
is linear.  This leads to a linear temperature profile.  The solutions
to the $J$ and $Q$ equations are unique only up to multiplicative
constants: they are scaled to match the total particle number in the
chain.

\begin{figure}
  \begin{center}
    \begin{small}
      \begin{tabular}{ccc}
        $D(x)$ &
        $\kappa(x)$ &
        $J(x)$ \\
        \resizebox{1.75in}{1.75in}{\includegraphics[bb=0in 0in 3in 2.5in]{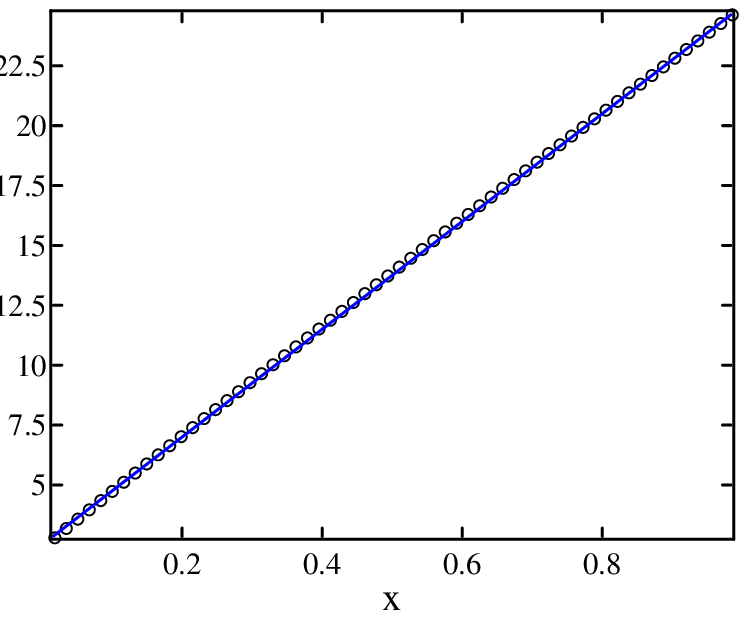}}&
        \resizebox{1.75in}{1.75in}{\includegraphics[bb=0in 0in 3in 2.5in]{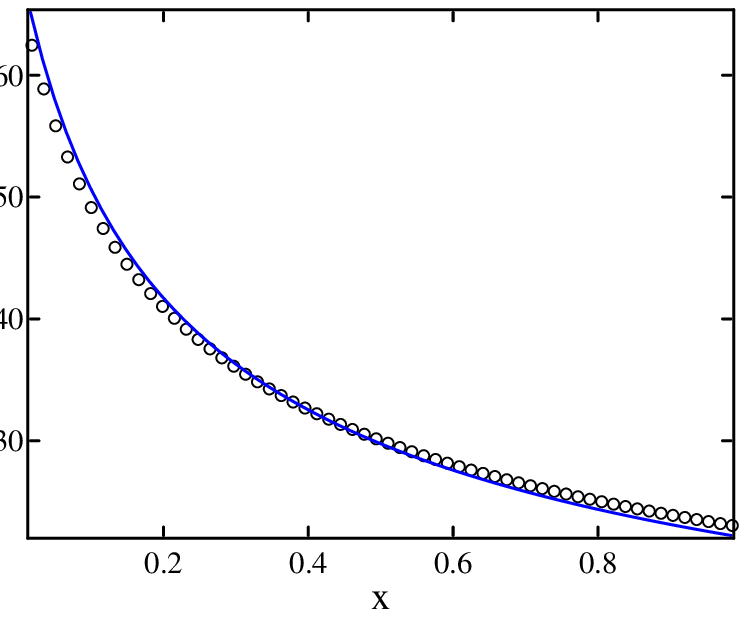}}&
        \resizebox{1.75in}{1.75in}{\includegraphics[bb=0in 0in 3in 2.5in]{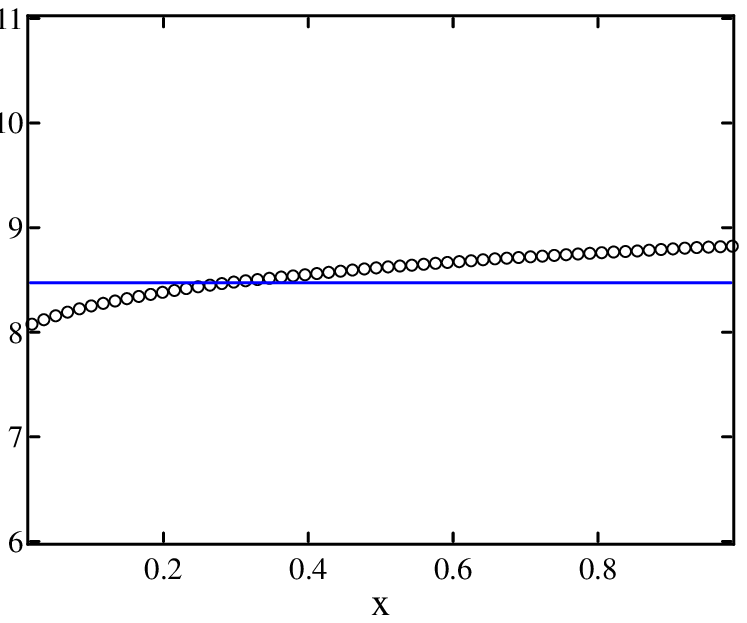}}\\

        \multicolumn{3}{c}{(a) Simulation and predicted profiles without
          bias correction}\\
      \end{tabular}\\[4ex]
      \resizebox{2.3in}{1.5in}{\includegraphics[bb=0in 0in 3.25in 2in]{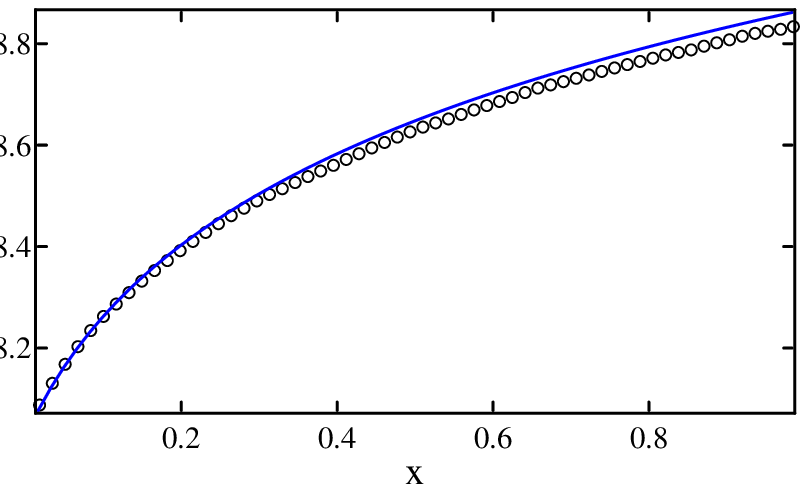}}\\
      (b) Bias-corrected $J$ profile
    \end{small}
  \end{center}
  \caption{Profiles for particle-conserving systems.  The system
    consists of 60 cells and contains 2000 particles.  Shown are
    simulation data (circles) and predicted profiles (solid lines).  The
    predictions in (a) are without bias correction; (b) uses the
    bias-corrected scheme.  Boundary conditions are $\tl=5$ and
    $\tr=50$.}
  \label{fig:11}
\end{figure}

\heading{Simulation results.}  Fig.~\ref{fig:11}(a) shows simulation
results for a chain with $N=60$ and $M=2000$, and boundary conditions
$T_L=5$ and $T_R=50$.  Shown are simulation data (dots) and predicted
profiles -- without any built-in bias correction.  As can be seen, the
agreement is very good except for the $J$ profile.

The discrepancy in the $J$ profile is, at first, a little puzzling: it
is somewhat larger than in similar non-particle-conserving chains, and
is not improved by doubling the total number of particles to $4000$.  At
the same time, estimates along the lines discussed in Sect.~3.1 show
that the amount of bias here ($\dr_{out} / J \leq 0.0025$) is relatively
small.  A possible explanation is that in the present setting, bias can
and does lead to a net gradient in $J$ across the length of the chain,
whereas in non-particle-conserving systems the values of $J$ at the two
ends are fixed by the baths.

Fig.~\ref{fig:11}(b) shows bias-corrected predictions for $J$ using the
same numerical values of the bias coefficients $\ar$, $\br$, etc., as in
Sect.~\ref{sect:memory}. Agreement with simulation data is good,
and lends further support to the idea that bias is a ``local''
phenomenon, governed mainly by the mean values of thermodynamic
quantities and their gradients.

\section*{Conclusions}
The main findings of this paper are:
\begin{enumerate}

\item {\em For a large class of Hamiltonian systems, macroscopic profiles
such as those for energy and particle count
can be quite accurately predicted in a simple way.} The scheme introduced
in \cite{ey} and applied to 1D chains is extended here to more general settings, including 2D systems on rectangular and hexagonal lattices 
driven by Dirichlet and other boundary conditions, even point sources. 
Strong agreement with simulation data is also obtained
for systems regulated by thermostats, and for
systems in which particle numbers are held constant.  

\item \emph{Memory effects resulting in directional bias of particle
trajectories can vary from negligible to significant depending on cell geometry.} 
Where such bias is weak, as is the case for a range
of examples, the prediction scheme above (which assumes zero bias)
gives accurate results. 
For ``bad" geometries, a scheme with built-in bias correction is shown
to be effective.

\item \emph{Porous cell boundaries (or large gaps) cause
finite-size effects to become more pronounced due to incomplete 
equilibration with baths or thermostats} but do not otherwise
appear to impact macroscopic profiles or LTE.
  
\end{enumerate}
A more detailed discussion of (ii) and (iii) is given in
Sect.~3.3.

\paragraph{Acknowledgement.}
We are grateful to the referee for many detailed and helpful
comments.


\end{document}